# A Mutual Information-Based Ensemble Kalman Filter

Tadashi TSUYUKI[1,2] and Shunji KOTSUKI[1,3,4]

[1] Center for Environmental Remote Sensing, Chiba University, Chiba, Japan

[2] Meteorological Research Institute, Japan Meteorological Agency, Tsukuba, Japan

[3] Institute for Advanced Academic Research, Chiba University, Chiba, Japan

[4] Research Institute of Disaster Medicine, Chiba University, Chiba, Japan

## Abstract

Ensemble Kalman filters (EnKFs) are widely used for data assimilation in geophysical systems. Among various implementations, the local ensemble transform Kalman filter (LETKF) has gained popularity because of its computational efficiency and suitability for large-scale applications. However, the deterministic EnKF such as the LETKF is known to be less robust than the stochastic EnKF in strongly nonlinear regimes. We generalize the LETKF such that it contains a stochastic term and includes the stochastic EnKF within it. We adaptively optimize the parameter that determines the weight of the stochastic term based on an identity of mutual information, which is satisfied by the Kalman filter in linear Gaussian systems. As the analysis perturbation equations of EnKFs are decomposed into a system of equations for modes that are uncorrelated with each other, the application of mutual information is easily achieved. The generalized LETKF thus optimized is named the mutual information-based ensemble Kalman filter (MI-EnKF). The MI-EnKF indirectly uses the third- and fourth-order moments of the forecast ensemble through entropy. To speed up calculations of entropy, we create a lookup table based on maximum entropy distributions. We conduct data assimilation experiments using the Lorenz-96 model to confirm the validity of the optimization method of MI-EnKF. When the observation operator is linear, the ML-EnKF shows the same analysis accuracy as the LETKF. When the observation operator is strongly nonlinear, the MI-EnKF is more accurate than both LETKF and stochastic EnKF regardless of ensemble size. Optimizing just the first mode can lead to significant improvements, but positive impacts of increasing the number of optimized modes are not observed unless the ensemble size is large. The optimized parameter values indicate that the optimal EnKF lies between the deterministic EnKF and the stochastic EnKF.

## 1. Introduction

Data assimilation is the methodology to estimate the state of a time-evolving complex system like the atmosphere and the ocean based on a numerical model and observational data of the system. It is an indispensable tool for research and forecasts in meteorology and oceanography. Ensemble Kalman filters (EnKFs; Evensen 1994) are widely used for data assimilation in geophysical systems. Among various implementations, the local ensemble transform Kalman filter (LETKF; Hunt et al. 2007) has gained popularity because of its computational efficiency and suitability for large-scale applications. However, Lawson and Hansen (2004) and Lei et al. (2010) showed that the deterministic EnKF such as the LETKF is less robust than the stochastic EnKF (Burgers et al. 1998, Houtekamer and Mitchell 1998) in strongly nonlinear regimes of data assimilation. Tsuyuki (2024) addressed this issue by revisiting EnKFs in a unified framework and proved that the analysis ensemble generated by the LETKF is uniform contractions of the forecast ensemble in observation space in each direction of the eigenvectors of the forecast error covariance matrix. This result implies that if the forecast ensemble is strongly non-Gaussian the analysis ensemble is also strongly non-Gaussian. Strong non-Gaussianity becomes more evident with an increase of ensemble size and tends to persist during data assimilation cycles performed at a high frequency. Consequently, the analysis accuracy of LETKF is degraded in strongly nonlinear regimes, because EnKFs are based on the Gaussian assumption. On the other hand, the analysis ensemble generated by the stochastic EnKF is more Gaussian due to Gaussian perturbations added to observations. A major problem with the stochastic EnKF is sampling noise introduced by perturbed observations. This EnKF is hereafter referred to as the perturbed observation ensemble Kalman filter (PO-EnKF).

Particle filters (PFs; Gordon et al. 1993, Kitagawa 1996) provide a fully nonlinear and non-Gaussian solution to the Bayesian filtering problem. However, their application to large geophysical systems is severely limited by the curse of dimensionality, which requires an exponentially increasing number of particles as the system dimension grows (Snyder et al., 2008). This limitation to applying PFs to high-dimensional systems is currently disappearing owing to the recent efforts of several investigators (van Leeuwen et al. 2019). The use of localization techniques within particle filtering frameworks has been proposed, leading to the development of the local particle filter (LPF; Penney and Miyoshi 2016, Poterjoy 2016). Subsequent studies demonstrated the feasibility of this approach in high-dimensional geophysical models (Poterjoy and Anderson 2016, Potthast et al. 2019, Kotsuki et al. 2022, Rojahn et al. 2023) and even in convective-scale numerical weather prediction experiments (Poterjoy et al., 2017). Although the LPF offers a promising framework for non-Gaussian data assimilation in high-dimensional systems, its practical application faces several major challenges (e.g., Farchi and Bocquet 2018); First, under a limited ensemble size, the LPF often fails to outperform EnKFs due to the insufficient sampling of localized state spaces. Second, the filter's accuracy is highly sensitive to empirical tuning parameters. Third, spatial discontinuity is introduced by independent local resampling, which often violates physical balances and generates spurious noise.

In recent decades, Information-theoretic approaches have increasingly been used to provide an alternative perspective on data assimilation by quantifying uncertainty reduction and information transfer during the analysis process. Early studies introduced entropy- and relative entropy-based measures to evaluate the information content of observations and demonstrated their applicability to observation impact assessment and adaptive observation design (Rodgers 2000, Xu 2007). These concepts were further extended to establish connections among information content measures and to support observation system design and optimization (Xu et al. 2008). Subsequently, information-theoretic metrics were incorporated into variational data assimilation frameworks to characterize the contribution of observations to uncertainty reduction (Singh et al. 2012). More recently, data assimilation itself has been interpreted as an information processing system, and measures based on entropy and mutual information have been proposed to quantify the efficiency of information extraction from observations (Nearing et al. 2018). These studies suggest that information-theoretic quantities provide complementary diagnostics beyond conventional estimation error-based evaluation criteria.

In this study, we generalize the LETKF such that it contains a stochastic term and includes both LETKF and PO-EnKF within it and adaptively optimize this EnKF using mutual information. More specifically, a parameter that determines the weight of the stochastic term is optimized such that an identity of mutual information holds. This identity is satisfied by the Kalman filter in linear Gaussian systems. Tsuyuki (2024) showed that the analysis perturbation equations of EnKFs are decomposed into a set of equations for one-dimensional systems that have no correlations with each other. The generalization of LETKF and the application of mutual information are based on this result. The generalized LETKF thus optimized is named the mutual information-based ensemble Kalman filter (MI-EnKF). The MI-EnKF indirectly uses the third- and fourth-order moments of the forecast ensemble through entropy. To speed up calculations of entropy, we create a lookup table based on maximum entropy distributions. To confirm the validity of the optimization method of MI-EnKF, we conduct data assimilation experiments using the Lorenz-96 model (Lorenz, 1996) with both linear and strongly nonlinear observation operators. As the MI-EnKF is based on the moments of the forecast ensemble, it can also handle the nonlinearity of numerical models.

The remainder of this paper is organized as follows. Section 2 introduces the revisit result of Tsuyuki (2024) and generalize the LETKF based on those results. Section 3 introduces mutual information and presents the optimization method using mutual information. Section 4 describes the design and results of data assimilation experiments with the Lorenz-96 model. A summary and discussion are mentioned in Section 5.

## 2. Generalization of LETKF

Tsuyuki (2024) revisited the deterministic EnKF and the stochastic EnKF in a unified framework. Since the result of that work provides a basis for the generalization of LETKF, we introduce that result first with additional remarks.

### *2.1. Revisit of EnKFs*

The analysis equation of EnKFs is based on an extension of Kalman filtering for a nonlinear observation operator $\mathcal{H}$:

$$\bar{\boldsymbol{x}}^a = \bar{\boldsymbol{x}}^f + \boldsymbol{K}(\boldsymbol{y}^o - \bar{\boldsymbol{y}}^f), \tag{1}$$

where $\bar{\boldsymbol{x}}^a$ and $\bar{\boldsymbol{x}}^f$ are the analysis ensemble average and forecast of the $n$-dimensional state variable $\boldsymbol{x}$, respectively, $\boldsymbol{y}^o$ is the observation of the $m$-dimensional observed variable $\boldsymbol{y}$, $\bar{\boldsymbol{y}}^f$ is the forecast ensemble average of $\boldsymbol{y}$, and $\boldsymbol{K}$ is the Kalman gain. $\boldsymbol{y}$ is defined by $\boldsymbol{y} = \mathcal{H}(\boldsymbol{x})$ using the observation operator $\mathcal{H}$. Let $N$ denote the ensemble size, and we introduce an $n \times N$ matrix $\boldsymbol{X}^f \coloneqq (\Delta\boldsymbol{x}^{f(1)}, \cdots, \Delta\boldsymbol{x}^{f(N)})$ and an $m \times N$ matrix $\boldsymbol{Y}^f \coloneqq (\Delta\boldsymbol{y}^{f(1)}, \cdots, \Delta\boldsymbol{y}^{f(N)})$, where $\Delta\boldsymbol{x}^{f(k)}$ and $\Delta\boldsymbol{y}^{f(k)}$ are the $k$th ensemble members of the forecast perturbations of the state variable and that of the observed variable, respectively.

The Kalman gain is obtained by adopting the minimum mean square error criterion:

$$\boldsymbol{K} = \frac{\boldsymbol{X}^f(\boldsymbol{Y}^f)^{\mathrm{T}}}{N-1}\left[\boldsymbol{R} + \frac{\boldsymbol{Y}^f(\boldsymbol{Y}^f)^{\mathrm{T}}}{N-1}\right]^{-1} = \frac{\boldsymbol{X}^f}{N-1}\left[\boldsymbol{I}_N + \frac{(\boldsymbol{Y}^f)^{\mathrm{T}}\boldsymbol{R}^{-1}\boldsymbol{Y}^f}{N-1}\right]^{-1}(\boldsymbol{Y}^f)^{\mathrm{T}}\boldsymbol{R}^{-1}, \tag{2}$$

where $\boldsymbol{R}$ is the observation error covariance matrix, $\boldsymbol{I}_N$ is the $N$-dimensional identity matrix, and the superscript $\mathrm{T}$ denotes the transpose of a vector and a matrix. The first formula in Eq. (2) is usually used in the PO-EnKF with the B-localization, whereas the second one is used in the LETKF with the R-localization (Greybush et al. 2011).

The PO-EnKF can be localized by using the Kalman gain of the LETKF with the R-localization, where perturbations added to observations are adjusted such that the ensemble averages of perturbations vanish, and the same perturbations are used for the same observations in neighboring local domains to avoid discontinuity. One of the methods for this is to assign a different initialization parameter for random number generation to each observation and to generate perturbations when observations are assimilated in local domains. This PO-EnKF is hereafter referred to as the localized perturbed observation ensemble Kalman filter (LPO-EnKF), which is as highly efficient as the LETKF through parallel computation.

We decompose the forecast and analysis ensemble perturbations of the LETKF and the LPO-EnKF into modes. First, we introduce the eigenvalue decomposition of a forecast error covariance matrix normalized by $\boldsymbol{R}$:

$$\frac{\boldsymbol{R}^{-1/2}\boldsymbol{Y}^f\left(\boldsymbol{R}^{-1/2}\boldsymbol{Y}^f\right)^{\mathrm{T}}}{N-1}=\boldsymbol{U}\boldsymbol{\Sigma}\boldsymbol{U}^{\mathrm{T}}, \tag{3}$$

where $\boldsymbol{\Sigma}$ is the diagonal matrix consisting of the eigenvalues in the descending order:

$$\boldsymbol{\Sigma}=\begin{cases}\mathrm{diag}\,[\sigma_1^2,\ \cdots,\sigma_{N-1}^2,\ 0,\ \cdots,\ 0] & (N\leq m)\\ \mathrm{diag}\,[\sigma_1^2,\ \cdots,\sigma_m^2] & (N>m)\end{cases}, \tag{4}$$

and $\boldsymbol{U}$ is the orthogonal matrix consisting of $m$-dimensional normalized eigenvectors $\{\boldsymbol{u}_i\}_{i=1}^m$. The eigenvalues $\sigma_i^2$ corresponds to the ratio of the forecast error variance to the observation error variance. We transform the forecast ensemble perturbations in observation space $\left\{\Delta\boldsymbol{y}^{f(k)}\right\}_{k=1}^{k=N}$ into $\left\{\left(\Delta\boldsymbol{z}_i^f\right)\right\}_{i=1}^m$ as follows:

$$\begin{pmatrix}\left(\Delta\boldsymbol{z}_1^f\right)^{\mathrm{T}}\\ \vdots\\ \left(\Delta\boldsymbol{z}_m^f\right)^{\mathrm{T}}\end{pmatrix}:=\boldsymbol{U}^{\mathrm{T}}\boldsymbol{R}^{-1/2}\boldsymbol{Y}^f. \tag{5}$$

where the transpose operation is introduced on the left-hand side such that $\Delta\boldsymbol{z}_i^f$ consists of the forecast ensemble members of the $i$th mode. The perturbations $\left\{\left(\Delta\boldsymbol{z}_i^f\right)\right\}_{i=1}^m$ satisfy the following equations:

$$\begin{cases}\left(\Delta\boldsymbol{z}_i^f\right)^{\mathrm{T}}\Delta\boldsymbol{z}_j^f=(N-1)\sigma_i^2\delta_{ij} & (i,\ j=1,\ \cdots,\ d)\\ \Delta\boldsymbol{z}_i^f=\boldsymbol{0} & (i=d+1,\ \cdots,\ m)\end{cases}, \tag{6}$$

where $\delta_{ij}$ is the Kronecker delta and $d:=\min(N-1,m)$. These equations indicate that the perturbations span a $d$-dimensional subspace and that they are uncorrelated with each other.

Although Tsuyuki (2024) did not mention this, we can directly derive $\left\{\left(\Delta\boldsymbol{z}_i^f\right)\right\}_{i=1}^d$ from the eigenvalue decomposition of another forecast error covariance matrix:

$$\frac{\left(\boldsymbol{Y}^f\right)^T\boldsymbol{R}^{-1}\boldsymbol{Y}^f}{N-1}=\boldsymbol{V}\boldsymbol{\Sigma}'\boldsymbol{V}^T, \tag{7}$$

where $\boldsymbol{\Sigma}'$ is the diagonal matrix consisting of the eigenvalues in the descending order:

$$\boldsymbol{\Sigma}'=\begin{cases}\mathrm{diag}\,[\sigma_1^2,\ \cdots,\sigma_{N-1}^2,\ 0] & (N\leq m)\\ \mathrm{diag}\,[\sigma_1^2,\ \cdots,\sigma_m^2,\ 0,\ \cdots,\ 0] & (N>m)\end{cases}, \tag{8}$$

and $\boldsymbol{V}$ is the orthogonal matrix consisting of $N$-dimensional normalized eigenvectors $\{\boldsymbol{v}_k\}_{k=1}^{N}$. This eigenvalue decomposition is conventionally used in the LETKF to compute the ensemble transformation matrix. The $i$th mode perturbation is given by

$$\Delta \boldsymbol{z}_i^f = \sqrt{N-1}\sigma_i \boldsymbol{v}_i. \tag{9}$$

Next, we derive analysis perturbation equations for the transformed analysis perturbations defined by

$$\begin{pmatrix} (\Delta \boldsymbol{z}_1^a)^{\mathrm{T}} \\ \vdots \\ (\Delta \boldsymbol{z}_m^a)^{\mathrm{T}} \end{pmatrix} \coloneqq \boldsymbol{U}^{\mathrm{T}} \boldsymbol{R}^{-1/2} \boldsymbol{Y}^a, \tag{10}$$

where $\boldsymbol{Y}^a \coloneqq \left(\Delta \boldsymbol{y}^{a(1)}, \cdots, \Delta \boldsymbol{y}^{a(N)}\right)$ and $\Delta \boldsymbol{y}^{a(k)}$ is the $k$th ensemble member of the analysis perturbations of the observed variable, and $\Delta \boldsymbol{z}_i^a$ consists of the analysis ensemble members of the $i$th mode. The analysis perturbation equations of the LETKF are given by

$$\Delta \boldsymbol{z}_i^a = \begin{cases} \dfrac{1}{\sqrt{1+\sigma_i^2}} \Delta \boldsymbol{z}_i^f & (i = 1, \cdots, d) \\ 0 & (i = d+1, \cdots, m) \end{cases}. \tag{11}$$

This equation reveals that the analysis perturbations of different modes have no correlation between them. The analysis ensemble perturbations in state space $\boldsymbol{X}^a \coloneqq \left(\Delta \boldsymbol{x}^{a(1)}, \cdots, \Delta \boldsymbol{x}^{a(N)}\right)$, where $\Delta \boldsymbol{x}^{a(k)}$ is the $k$ th ensemble member of the forecast perturbations of the state variable, is computed by

$$\boldsymbol{X}^a = \boldsymbol{X}^f \boldsymbol{T}, \tag{12}$$

The ensemble transformation matrix $\boldsymbol{T}$ is expressed as

$$\boldsymbol{T} \coloneqq \left[\boldsymbol{I}_N + \frac{\left(\boldsymbol{Y}^f\right)^T \boldsymbol{R}^{-1} \boldsymbol{Y}^f}{N-1}\right]^{-1/2} = \boldsymbol{I}_N - \frac{1}{N-1} \sum_{i=1}^{d} \frac{1}{\sigma_i^2} \left(1 - \frac{1}{\sqrt{1+\sigma_i^2}}\right) \Delta \boldsymbol{z}_i^f \left(\Delta \boldsymbol{z}_i^f\right)^T. \tag{13}$$

For the LPO-EnKF, we introduce Gaussian perturbations to observations $\left\{\boldsymbol{\varepsilon}^{o(k)}\right\}_{k=1}^{N}$, which are generated by

$$\boldsymbol{\varepsilon}^{o(k)} \coloneqq \boldsymbol{\varepsilon}^{o(k)*} - \frac{1}{N} \sum_{k=1}^{N} \boldsymbol{\varepsilon}^{o(k)*}, \qquad \boldsymbol{\varepsilon}^{o(k)*} \sim \mathcal{N}(\boldsymbol{0}, \boldsymbol{R}), \tag{14}$$

where $\mathcal{N}(\boldsymbol{0}, \boldsymbol{R})$ denotes the Gaussian distribution with mean $\boldsymbol{0}$ and covariance $\boldsymbol{R}$. We introduce the transformed perturbations $\{\boldsymbol{f}_i^o\}_{i=1}^{m}$ defined by

$$\begin{pmatrix} (\boldsymbol{f}_1^o)^{\mathrm{T}} \\ \vdots \\ (\boldsymbol{f}_m^o)^{\mathrm{T}} \end{pmatrix} = \boldsymbol{U}^{\mathrm{T}} \boldsymbol{R}^{-1/2} \boldsymbol{E}^o \tag{15}$$

where $\boldsymbol{E}^o \coloneqq \left(\boldsymbol{\varepsilon}^{o(1)},\ \cdots, \boldsymbol{\varepsilon}^{o(N)}\right)$. They satisfy the following orthogonal condition:

$$\langle (\boldsymbol{f}_i^o)^{\mathrm{T}} \boldsymbol{f}_j^o \rangle \coloneqq (N-1)\, \delta_{ij} \qquad (i, j = 1,\ \cdots,\ m), \tag{16}$$

where a pair of brackets denotes the expectation operator. When Eq. (7) is used instead of Eq. (3) for the eigenvalue decomposition, $\boldsymbol{f}_i^o$ can be computed by

$$\boldsymbol{f}_i^o = (\boldsymbol{E}^o)^{\mathrm{T}} \boldsymbol{R}^{-1/2} \boldsymbol{u}_i = (\boldsymbol{E}^o)^{\mathrm{T}} \frac{\boldsymbol{R}^{-1} \boldsymbol{Y}^f \boldsymbol{v}_i}{\| \boldsymbol{R}^{-1/2} \boldsymbol{Y}^f \boldsymbol{v}_i \|}. \tag{17}$$

The analysis perturbation equations of the LPO-EnKF are given by

$$\Delta \boldsymbol{z}_i^a = \begin{cases} \dfrac{1}{1+\sigma_i^2} \Delta \boldsymbol{z}_i^f + \dfrac{\sigma_i^2}{1+\sigma_i^2} \boldsymbol{f}_i^o & (i = 1,\ \cdots,\ d) \\ 0 & (i = d+1,\ \cdots,\ m) \end{cases}. \tag{18}$$

Taking expectation values of $(\Delta \boldsymbol{z}_i^a)^{\mathrm{T}} \Delta \boldsymbol{z}_j^a$ and using Eqs. (6) and (16), we observe that the analysis perturbations of different modes of LPO-EnKF have no correlation between them. It may be desirable to remove correlation with $\Delta \boldsymbol{z}_i^f$ from $\boldsymbol{f}_i^o$ and to adjust the resulting perturbation such that its variance is left unchanged. This procedure is adopted in the data assimilation experiments of this study.

In summary, the analysis perturbation equations of LETKF and LPO-EnKF are decomposed into a set of equations for one-dimensional systems that have no correlations with each other. Although Tsuyuki (2024) derived Eqs. (11) and (18) under the assumption that the observation operator is linear, these equations are also valid even if it is nonlinear. We can show this using the joint state-observation space method (Anderson, 2001), in which $\boldsymbol{Y}^a = \boldsymbol{Y}^f \boldsymbol{T} \neq \mathcal{H}(\boldsymbol{X}^a) = \mathcal{H}\left(\boldsymbol{X}^f \boldsymbol{T}\right)$ while $\boldsymbol{Y}^f = \mathcal{H}\left(\boldsymbol{X}^f\right)$ for the LETKF as an example, where the operation of $\mathcal{H}$ is performed on each ensemble member although it is a rough representation.

### *2.2. Generalized LETKF*

We formulate a generalized LETKF based on the revisit result. We notice that Eqs. (9) and (10) can be generalized as

$$\Delta \boldsymbol{z}_i^a(w_i) = \begin{cases} \dfrac{w_i}{\sqrt{1+\sigma_i^2}} \Delta \boldsymbol{z}_i^f + \sqrt{\dfrac{(1-{w_i}^2)\sigma_i^2}{1+\sigma_i^2}} \boldsymbol{f}_i^o & (i = 1,\ \cdots,\ d) \\ 0 & (i = d+1,\ \cdots,\ m) \end{cases}, \tag{19}$$

where $\{w_i\}_{i=1}^d$ are the weight parameters satisfying $0 \le w_i \le 1$. This equation is reduced to Eq. (11) if $w_i$ is set to 1 and reduced to Eq. (18) if $w_i$ is set to $1/\sqrt{1+\sigma_i^2}$. The following equation is easily derived from Eqs. (6), (16), and (19):

$$\frac{\left\langle \left(\Delta \boldsymbol{z}_i^a(w_i)\right)^{\mathrm{T}} \Delta \boldsymbol{z}_j^a(w_i) \right\rangle}{N-1} = \frac{\sigma_i^2}{1+\sigma_i^2} \delta_{ij} \qquad (i, j = 1, \cdots, d)\,. \tag{20}$$

This equation indicates that the variances of analysis perturbations do not depend on the weight parameters, and that the analysis perturbations of different modes have no correlations between them. We therefore see that Eq. (19) is a correct generalization of Eqs. (11) and (18). Note that this generalization can be obtained by adding Gaussian perturbations to the right-hand side of Eq. (11) and rescale $\Delta \boldsymbol{z}_i^a(w_i)$ such that Eq. (20) holds. Using the revisit result, however, we can establish the relationship between $\{\boldsymbol{f}_i^o\}_{i=1}^d$ and $\{\boldsymbol{\varepsilon}^{o(k)}\}_{k=1}^N$ of Eq. (15) and avoid discontinuity between adjacent local domains by employing the procedure mentioned in the third paragraph of Subsection 2.1. This procedure is adopted in the data assimilation experiments of this study.

An inspection of Eqs. (11) and (13) suggests the following generalization of the ensemble transformation matrix:

$$\widetilde{\boldsymbol{T}} \coloneqq \boldsymbol{I}_N - \frac{1}{N-1} \sum_{i=1}^{d} \frac{1}{\sigma_i^2} \Delta \boldsymbol{z}_i^f \left( \Delta \boldsymbol{z}_i^f - \Delta \boldsymbol{z}_i^a(w_i) \right)^{\mathrm{T}}, \tag{21}$$

and the analysis ensemble perturbations in state space are computed by

$$\boldsymbol{X}^a = \boldsymbol{X}^f \widetilde{\boldsymbol{T}}. \tag{22}$$

Substitution of Eq. (19) into Eq. (21) yields

$$\widetilde{\boldsymbol{T}} = \boldsymbol{I}_N - \frac{1}{N-1} \sum_{i=1}^{d} \frac{1}{\sigma_i^2} \left( 1 - \frac{w_i}{\sqrt{1+\sigma_i^2}} \right) \Delta \boldsymbol{z}_i^f \left( \Delta \boldsymbol{z}_i^f \right)^{\mathrm{T}} + \sum_{i=1}^{d} \frac{1}{\sigma_i} \sqrt{\frac{1-{w_i}^2}{1+\sigma_i^2}} \Delta \boldsymbol{z}_i^f (\boldsymbol{f}_i^o)^{\mathrm{T}}. \tag{23}$$

The expectation value of $\widetilde{\boldsymbol{T}}\widetilde{\boldsymbol{T}}^{\mathrm{T}}$ is computed by using Eq. (16) as

$$\langle \widetilde{\boldsymbol{T}}\widetilde{\boldsymbol{T}}^{T} \rangle = I_N - \frac{1}{N-1} \sum_{i=1}^{d} \frac{1}{\sigma_i^2} \left( 1 - \frac{1}{1+\sigma_i^2} \right) \Delta \boldsymbol{z}_i^f \left( \Delta \boldsymbol{z}_i^f \right)^{\mathrm{T}} = \boldsymbol{T}^2. \tag{24}$$

Consequently, the expectation value of the analysis error covariance matrix $\boldsymbol{P}^a$ is given by

$$\langle \boldsymbol{P}^a \rangle = \frac{\langle \boldsymbol{X}^a (\boldsymbol{X}^a)^{\mathrm{T}} \rangle}{N-1} = \frac{\boldsymbol{X}^f \boldsymbol{T}^2 \left( \boldsymbol{X}^f \right)^{\mathrm{T}}}{N-1}. \tag{25}$$

The matrix on the rightmost side of this equation is the analysis error covariance matrix of LETKF, so that we can confirm that Eqs. (21) and (22) generate correct analysis ensemble perturbations. There may be a concern that when eigenvalue $\sigma_i^2$ approaches 0, Eq. (23) may diverge. However, it is easily shown that divergence does not occur if $w_i = 1 + O(\sigma_i^2)$ when $\sigma_i^2 \ll 1$. This condition is satisfied by both LETKF ($w_i = 1$) and LPO-EnKF ($w_i = 1/\sqrt{1+\sigma_i^2}$). As will be mentioned in Subsection 3.2, we optimize the modes with large eigenvalues only.

We need the analysis ensemble average. It may be natural from Eq. (22) to use the following Kalman gain:

$$\widetilde{\boldsymbol{K}} = \frac{1}{N-1}\boldsymbol{X}^f\widetilde{\boldsymbol{T}}\widetilde{\boldsymbol{T}}^{\mathrm{T}}\left(\boldsymbol{Y}^f\right)^{\mathrm{T}}\boldsymbol{R}^{-1}\,. \tag{26}$$

However, this Kalman gain contains $\{\boldsymbol{f}_i^o\}_{i=1}^d$ that introduce sampling noise. We therefore take the expectation of Eq. (26) and obtain the same Kalman gain as that of LETKF. In summary, the analysis ensemble average of the generalized LETKF is the same as that of LETKF, whereas the analysis ensemble perturbations are generated by Eqs. (21) and (22) along with Eq. (19). If the weight parameters are appropriately optimized, we can expect that the generalized LETKF will generate better analysis than the LETKF during data assimilation cycles.

## 3. Optimization using mutual information

The generalized LETKF presented in the previous section contains the weight parameters. We apply information theory to adaptively optimize these parameters using mutual information. As the analysis perturbation equations are decomposed into a set of equations for one-dimensional systems, the application of information theory is easily achieved.

### *3.1. Identity of mutual information*

One of the fundamental concepts of information theory is mutual information (e.g., Cover and Thomas 2006). Let $p(\boldsymbol{x},\,\boldsymbol{y})$ be the joint probability density function (PDF) of two random variables $X, Y$, and let $p(\boldsymbol{x})$ and $p(\boldsymbol{y})$ be the marginal PDFs of $X$ and $Y$, respectively. The mutual information between them is defined by

$$I[X,\,Y] \coloneqq \int_{-\infty}^{\infty}\int_{-\infty}^{\infty} p(\boldsymbol{x},\,\boldsymbol{y})\,\log\frac{p(\boldsymbol{x},\,\boldsymbol{y})}{p(\boldsymbol{x})\,p(\boldsymbol{y})}\,d\boldsymbol{x}d\boldsymbol{y}\,. \tag{27}$$

This is the Kullback-Leibler divergence between $p(\boldsymbol{x},\,\boldsymbol{y})$ and $p(\boldsymbol{x})\,p(\boldsymbol{y})$, We can rewrite this definition using entropy $H[\cdot]$ and conditional entropy $H[\cdot\mid\cdot]$ as

$$I[X,\,Y] = H[X] - H[X|Y] = H[Y] - H[Y|X]. \tag{28}$$

These entropies are computed from PDFs as

$$H[X] \coloneqq -\int_{-\infty}^{\infty} p(\boldsymbol{x})\,\log p(\boldsymbol{x})\,d\boldsymbol{x}\,, \tag{29}$$

$$H[X|Y] \coloneqq -\int_{-\infty}^{\infty}\int_{-\infty}^{\infty} p(\boldsymbol{x},\,\boldsymbol{y})\,\log p(\boldsymbol{x}|\,\boldsymbol{y})\,d\boldsymbol{x}d\boldsymbol{y} = \int_{-\infty}^{\infty} p(\boldsymbol{y})\,H[X|Y=\boldsymbol{y}]\,d\boldsymbol{y}\,, \tag{30}$$

where $p(\boldsymbol{x}|\,\boldsymbol{y})$ is the conditional PDF of $X$ with given $Y$. Since entropy is a measure of uncertainty of a random variable, $I[X,\,Y]$ gives a reduction in the uncertainty of $X$ due to the knowledge of $Y$ and vice versa. The second equality in Eq. (28) is hereafter referred to as the identity of mutual information.

In the context of data assimilation, we can regard $X$ and $Y$ as the state variable and the observational data, respectively. $H[X]$ is the entropy of the forecast ensemble, and $H[X|Y]$ is that of the analysis ensemble. The analysis ensemble of EnKFs does not depend on observations. Consequently, we do not need $p(\boldsymbol{y})$ in computing $H[X|Y]$ in Eq. (30), and we can replace $H[X|Y]$ by the entropy of an analysis ensemble for specific observations $\boldsymbol{y}^o$:

$$H_{\mathrm{EnKF}}[X|Y] = H_{\mathrm{EnKF}}[X|Y = \boldsymbol{y}^o]\,. \tag{31}$$

This result simplifies the calculation of $H[X|Y]$. Furthermore, in Kalman filtering, the observation error is usually assumed to be additive and Gaussian. Therefore, $H[Y|X]$ is equal to the entropy of $\mathcal{N}(\mathbf{0}, \boldsymbol{R})$. The computation of $H[Y]$ needs $p(\boldsymbol{y})$, which are computed from the likelihood function and the forecast ensemble.

We can show that in linear Gaussian systems the Kalman filter is derived by maximizing mutual information and that it satisfies the identity of mutual information in Eq. (28) (see Appendix 1). However, these two statements are not valid in non-Gaussian systems. We examine the latter issue by taking an example from Subsection 3a of Lawson and Hansen (2004). It is a one-dimensional system with a prior PDF given by

$$p(x) = \frac{1}{2\sqrt{2\pi}}\left\{\exp\left[-\frac{(x-4)^2}{2}\right] + \exp\left[-\frac{(x+4)^2}{2}\right]\right\}, \tag{32}$$

and a likelihood function given by

$$p(y^o|x) = \frac{1}{\sqrt{2\pi}\sigma^o}\exp\left[-\frac{(y^o - x)^2}{2(\sigma^o)^2}\right]. \tag{33}$$

The forecast error standard deviation $\sigma^f$ is calculated as $\sqrt{17}$, and the observation error standard deviation $\sigma^o$ is set to $\sigma^f/2$. The analysis PDFs of ETKF (LETKF without localization) and PO-EnKF are derived from Eqs. (11) and (18) with $n = m = 1$ and $N \to \infty$ as follows:

$$p_{\mathrm{ETKF}}(x|y^o) = \sqrt{1+\sigma^2}\, p\left(\sqrt{1+\sigma^2}(x - x^a_{\mathrm{KF}})\right), \tag{34}$$

$$p_{\mathrm{PO-EnKF}}(x|y^o) = \frac{(1+\sigma^2)^2}{\sigma^2}\int_{-\infty}^{\infty} p\big((1+\sigma^2)(x - y - x^a_{\mathrm{KF}})\big)\, p\left(\frac{1+\sigma^2}{\sigma^2}y\middle|0\right) dy, \tag{35}$$

where $\sigma = \sigma^f/\sigma^o = 2$ and $x^a_{\mathrm{KF}} := \sigma^2 y^o/(1+\sigma^2)$. Equations (32) – (35) are plotted in Fig. 1 for $y^o = 3.5$, along with the posterior PDF $p(x|y^o)$. Since the Kalman filter in non-Gaussian systems is not optimal, its analysis $x^a_{\mathrm{KF}} = 2.8$ is different from the mean of $p(x|y^o)$, which is calculated as 3.874. The analysis PDFs of the two EnKFs are quite different from each other, although the means and variances are the same. As Eq. (34) suggests, the analysis PDF of ETKF is a uniform contraction of the prior PDF.

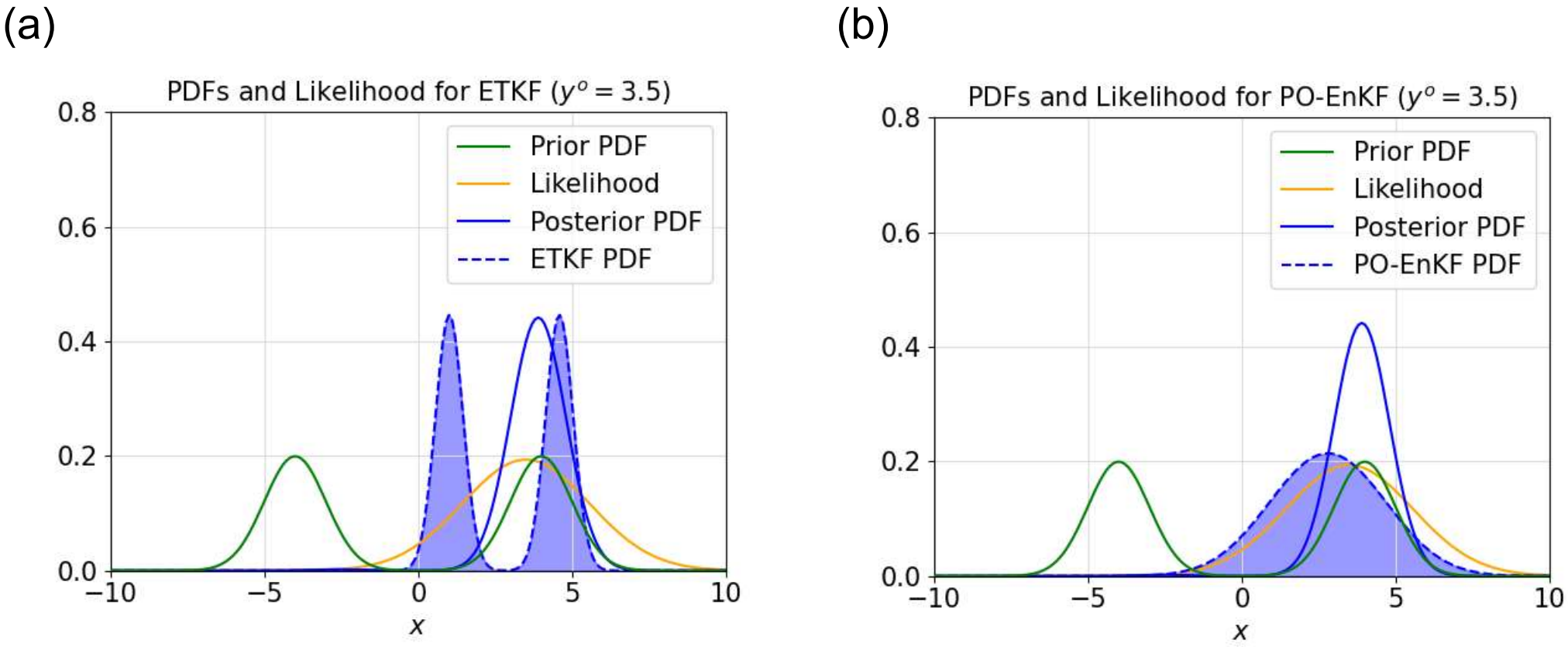

Fig. 1. Prior PDF (green line), likelihood function (orange line), posterior PDF (solid blue line), and analysis PDF (dashed blue line with filling) of (a) ETKF and (b) PO-EnKF for an example of Lawson and Hansen (2004). The observation $y^o$ is set to 3.5.

The values of each entropy in Eq. (28) of this example are presented in the "Exact" row in Table 1. These are obtained by numerical integration. It is confirmed that the identity of mutual information holds even if the prior PDF is strongly non-Gaussian. Note that $H[X|Y]$ is different from $H[X|Y = y^o]$, which is calculated as 1.343. On the other hand, the entropies of analysis PDFs of the two EnKFs do not satisfy the identity of mutual information. $H_{\mathrm{ETKF}}[X|Y]$ is smaller than $H[X|Y]$, indicating that irrelevant information has been mixed in. $H_{\mathrm{PO-En}}$ $[X|Y]$ is larger than $H[X|Y]$, indicating that relevant information has been lost. This result suggests that we could optimize the weight parameters of the generalized LETKF by requiring that the identity of mutual information holds. This is the key idea of this study.

Table 1. Entropies of a one-dimensional example of Lawson and Hansen (2004) shown in Fig. 1. $H[X|Y]$ in the "Table lookup" row is not a table lookup result, but it is computed using $H[X|Y] = H[X] - H[Y] + H[Y|X]$. The normalized third- and fourth-order moments used in table lookup are also presented. For table lookup, see Subsection 3.3.

| | $H[X]$ | $H[X\|Y]$ | $H[Y]$ | $H[Y\|X]$ | $H_{\mathrm{ETKF}}[X\|Y]$ | $H_{\mathrm{PO-EnKF}}[X\|Y]$ |
|---|---|---|---|---|---|---|
| Exact | 2.111 | 1.417 | 2.837 | 2.142 | 1.307 | 2.030 |
| Table lookup | 2.167 | 1.464 | 2.845 | 2.142 | 1.363 | 2.031 |
| $\widetilde{m}_3$ | 0 | - | 0 | 0 | 0 | 0 |
| $\widetilde{m}_4$ | 1.228 | - | 1.866 | 3 | 1.228 | 2.928 |

### *3.2. Method of optimization*

We adaptively optimize the weight parameters of the generalized LETKF for the first $d_c\ (\ll d)$ modes to satisfy the identity of mutual information in Eq. (28). Those of the remaining modes are set to 1. Since Eq. (19) consists of equations for one-dimensional systems, we can easily compute entropy. We solve the following equations for $\{w_i\}_{i=1}^{d_c}$ using Newton's method subject to the constraint of $0 \le w_i \le 1$:

$$H[\Delta \boldsymbol{z}_i^a(w_i)] = H\left[\Delta \boldsymbol{z}_i^f\right] - H\left[\Delta \boldsymbol{z}_i^f + \boldsymbol{f}_i^o\right] + H[\boldsymbol{f}_i^o] \qquad (\, i = 1, \cdots,\, d_c)\,, \tag{36}$$

Each entropy from left to right corresponds to $H[X|Y]$, $H[X]$, $H[Y]$, and $H[Y|X]$ in Eq. (28), respectively; the vector consisting of ensemble members is used as the argument of entropy instead of the random variable. The first guess of $\{w_i\}_{i=1}^{d_c}$ is set to 1; in other words, we start with the LETKF in iterative calculations. Equation (19) indicates that more reduction in $w_i$ introduces more Gaussian perturbations to $\Delta \boldsymbol{z}_i^a(w_i)$. Therefore, we can expect that $H[\Delta \boldsymbol{z}_i^a(w_i)]$ is a decreasing function of $w_i$. Then, we solve Eq. (36) only if $H[\Delta \boldsymbol{z}_i^a(1)]$ is smaller than the right-hand side of this equation. If this condition is not satisfied, we give up the optimization and adopt the LETKF. Note that each equation in Eq. (36) can be solved in parallel.

To solve Eq. (36), we need to compute entropy from an ensemble. We estimate entropy from the first four moments of the ensemble based on the maximum entropy method (e.g., Cover and Thomas 2006). Since the first-order moment $m_1$ of each set of ensemble members in Eq. (36) vanishes, we need the remaining three moments. In the following equations up to Eq. (45), the subscript $i$ that represents the mode number is omitted to avoid becoming cumbersome. For the forecast perturbations $\Delta \boldsymbol{z}^f$, the three moments are calculated as

$$m_2^f = \frac{1}{N-1}\sum_{k=1}^{N}\left(\Delta z^{f(k)}\right)^2 = \sigma^2, \quad m_3^f = \frac{N}{(N-1)(N-2)}\sum_{k=1}^{N}\left(\Delta z^{f(k)}\right)^3,$$
$$m_4^f = \frac{N(N+1)}{(N-1)(N-2)(N-3)}\sum_{k=1}^{N}\left(\Delta z^{f(k)}\right)^4 - \frac{3(3N-5)}{(N-2)(N-3)}\sigma^4. \tag{37}$$

The second equality of the first equation is derived from Eq. (6). As the other perturbations in Eq. (36) contain $\boldsymbol{f}_i^o$, we take the expectation values of moments and use $\{\langle m_l \rangle\}_{l=1}^4$ instead of $\{m_l\}_{l=1}^4$ to suppress the adverse effect of sampling noise. For the sum of forecast and observation perturbations $\Delta \boldsymbol{z}^f + \boldsymbol{f}^o$,

$$\langle m_2^s \rangle = 1 + \sigma^2, \qquad \langle m_3^s \rangle = m_3^f, \qquad \langle m_4^s \rangle = m_4^f + 6\sigma^2 + 3. \tag{38}$$

For the analysis perturbations $\Delta \boldsymbol{z}^a(w)$,

$$\langle m_2^a \rangle = \frac{\sigma^2}{1+\sigma^2}, \quad \langle m_3^a \rangle = \frac{w^3}{(1+\sigma^2)^{3/2}} m_3^f, \quad \langle m_4^a \rangle = \frac{w^4}{(1+\sigma^2)^2} m_4^f + \frac{3(1-w^4)}{(1+\sigma^2)^2}\sigma^4. \tag{39}$$

For the observation perturbations $\boldsymbol{f}^o$,

$$\langle m_2^o \rangle = 1, \qquad \langle m_3^o \rangle = 0, \qquad \langle m_4^o \rangle = 3. \tag{40}$$

We introduce the normalized moment $\widetilde{m}_l \coloneqq m_l / {m_2}^{l/2}$ for convenience. Then, $\widetilde{m}_1 = 0$ and $\widetilde{m}_2 = 1$. Note that $\widetilde{m}_3$ is skewness and $\widetilde{m}_4$ is kurtosis. Equations (38) and (39) can be rewritten as

$$\langle \widetilde{m}_3^s \rangle = \left(\frac{\sigma}{1+\sigma^2}\right)^{3/2} \widetilde{m}_3^f, \qquad \langle \widetilde{m}_4^s \rangle - 3 = \left(\frac{\sigma^2}{1+\sigma^2}\right)^2 \left(\widetilde{m}_4^f - 3\right), \tag{41}$$

$$\langle \widetilde{m}_3^a \rangle = w^3 \widetilde{m}_3^f, \qquad \langle \widetilde{m}_4^a \rangle - 3 = w^4 \left(\widetilde{m}_4^f - 3\right), \tag{42}$$

respectively. These equations suggest that the histograms of $\Delta \boldsymbol{z}^f + \boldsymbol{f}^o$ and $\Delta \boldsymbol{z}^a(w)$ are more Gaussian than that of $\Delta \boldsymbol{z}^f$. Entropy $H[X]$ is converted to normalized entropy $\widetilde{H}[X]$:

$$\widetilde{H}[X] \coloneqq H[X] - \frac{1}{2} \log m_2. \tag{43}$$

As entropy does not depend on the sign of $\widetilde{m}_3$, $\widetilde{H}[X]$ is a function of $|\widetilde{m}_3|$ and $\widetilde{m}_4$ only. Equation (36) is rewritten as

$$\widetilde{H}[\Delta \boldsymbol{z}^a(w)] = \widetilde{H}\left[\Delta \boldsymbol{z}^f\right] - \widetilde{H}\left[\Delta \boldsymbol{z}^f + \boldsymbol{f}^o\right] + \left(\log \sqrt{2\pi} + \frac{1}{2}\right). \tag{44}$$

We solve this equation for $w$ using Newton's method with Eqs. (37), (41), and (42) only if $\widetilde{H}[\Delta \boldsymbol{z}^a(1)]$ is smaller than the right-hand side. The derivative of $\widetilde{H}[\Delta \boldsymbol{z}^a(w)]$ with respect to $w$ is given by

$$\frac{d\widetilde{H}[\Delta \boldsymbol{z}^a(w)]}{dw} = \frac{1}{w}\left[3 \frac{\partial \widetilde{H}[\Delta \boldsymbol{z}^a(w)]}{\partial |\langle \widetilde{m}_3^a \rangle|} |\langle \widetilde{m}_3^a \rangle| + 4 \frac{\partial \widetilde{H}[\Delta \boldsymbol{z}^a(w)]}{\partial \langle \widetilde{m}_4^a \rangle} (\langle \widetilde{m}_4^a \rangle - 3)\right], \tag{45}$$

where Eq. (42) is used.

### *3.3. Lookup table for entropy*

As mentioned in the previous subsection, we estimate entropy from the first four moments of an ensemble using the maximum entropy method. We create a lookup table for entropy to avoid computing entropy every time. According to Cover and Thomas (2006), the maximum entropy distribution that satisfies the constraint that the first four moments $\{m_l\}_{l=1}^4$ are given is

$$p(x) = \exp\left(-\sum_{l=0}^{4} \lambda_l x^l\right). \tag{46}$$

The entropy of maximum entropy distribution is given by

$$H[X] = \sum_{l=0}^{4} \lambda_l \, m_l, \tag{47}$$

where $m_0 = 1$ is introduced for convenience. The parameters $\{\lambda_k\}_{k=0}^4$ satisfy the following equations:

$$\int_{-\infty}^{\infty} x^l \exp\left(-\sum_{k=0}^{4} \lambda_k x^k\right) dx = m_l \qquad (l = 0, \cdots, 4). \tag{48}$$

This set of equations is solved for $\{\lambda_k\}_{k=0}^4$ using Newton's method, and substitution of the solution into Eq. (47) yields entropy. If we use normalized moments $\{\widetilde{m}_l\}_{l=1}^4$ in solving Eq. (48), Eq. (47) gives normalized entropy $\widetilde{H}[X]$. The preparation process of the lookup table is briefly described in Appendix 2.

Figure 2a displays a lookup table obtained by the maximum entropy method. The grid intervals of $|\widetilde{m}_3|$ and $\widetilde{m}_4$ are both set to 0.02. Grid points where Newton's method did not converge are left blank. Normalized entropy takes the maximum value $\log\sqrt{2\pi} + 1/2$ at $(|\widetilde{m}_3|, \widetilde{m}_4) = (0, 3)$, where the Gaussian distribution is located. As proved in Appendix 3, $|\widetilde{m}_3|$ and $\widetilde{m}_4$ satisfy an inequality $\widetilde{m}_4 \geq |\widetilde{m}_3|^2 + 1$, so that there is a blank area on the right of the graph $\widetilde{m}_4 = |\widetilde{m}_3|^2 + 1$ (lower dashed line). There are other blank areas on the upper left. This is because the maximum entropy distribution Eq. (46) cannot represent heavy-tailed distributions such as Student's $t$ distribution. Although Newton's method sporadically converged in those areas, there is no guarantee that correct values are computed. Then, we fill in the left of an empirical graph $\widetilde{m}_4 = 5\,|\widetilde{m}_3|^{2.5} + 3$ (upper dashed line) with the entropy of mixture distributions of which PDFs are constructed by linear interpolation of known PDFs on the $(|\widetilde{m}_3|, \widetilde{m}_4)$ plane.

Figure 2b displays the lookup table adopted in this study. We estimate $\widetilde{H}[X]$ from the values of $|\widetilde{m}_3|$ and $\widetilde{m}_4$ by using this table. Equations (41) and (42) imply that $(|\langle\widetilde{m}_3^s\rangle|, \langle\widetilde{m}_4^s\rangle)$ and $(|\langle\widetilde{m}_3^a\rangle|, \langle\widetilde{m}_4^a\rangle)$ are located closer to (0, 3) than $\left(|\widetilde{m}_3^f|, \widetilde{m}_4^f\right)$ on the $(|\widetilde{m}_3|, \widetilde{m}_4)$ plane, so that we can estimate all entropies in Eq. (42) unless $\left(|\widetilde{m}_3^f|, \widetilde{m}_4^f\right)$ falls into the blank area of Fig. 2b. Otherwise, we give up the optimization and adopt the LETKF. However, there is a problem with this lookup table. Dots in the table represent the grid points where the quantity inside the square bracket in Eq. (45) is positive; in other words, the derivative of $H[\Delta \boldsymbol{z}_i^a(w)]$ with respect to $w$ is positive. As this is the opposite to what we expect, the calculated values of entropy at those grid points are questionable. Note that there are no such grid points on the right of $\widetilde{m}_4 = 5\,|\widetilde{m}_3|^{2.5} + 3$ where Newton's method converged. Fortunately, $\left(|\widetilde{m}_3^f|, \widetilde{m}_4^f\right)$ rarely fell into the dotted areas in the data assimilation experiments of this study.

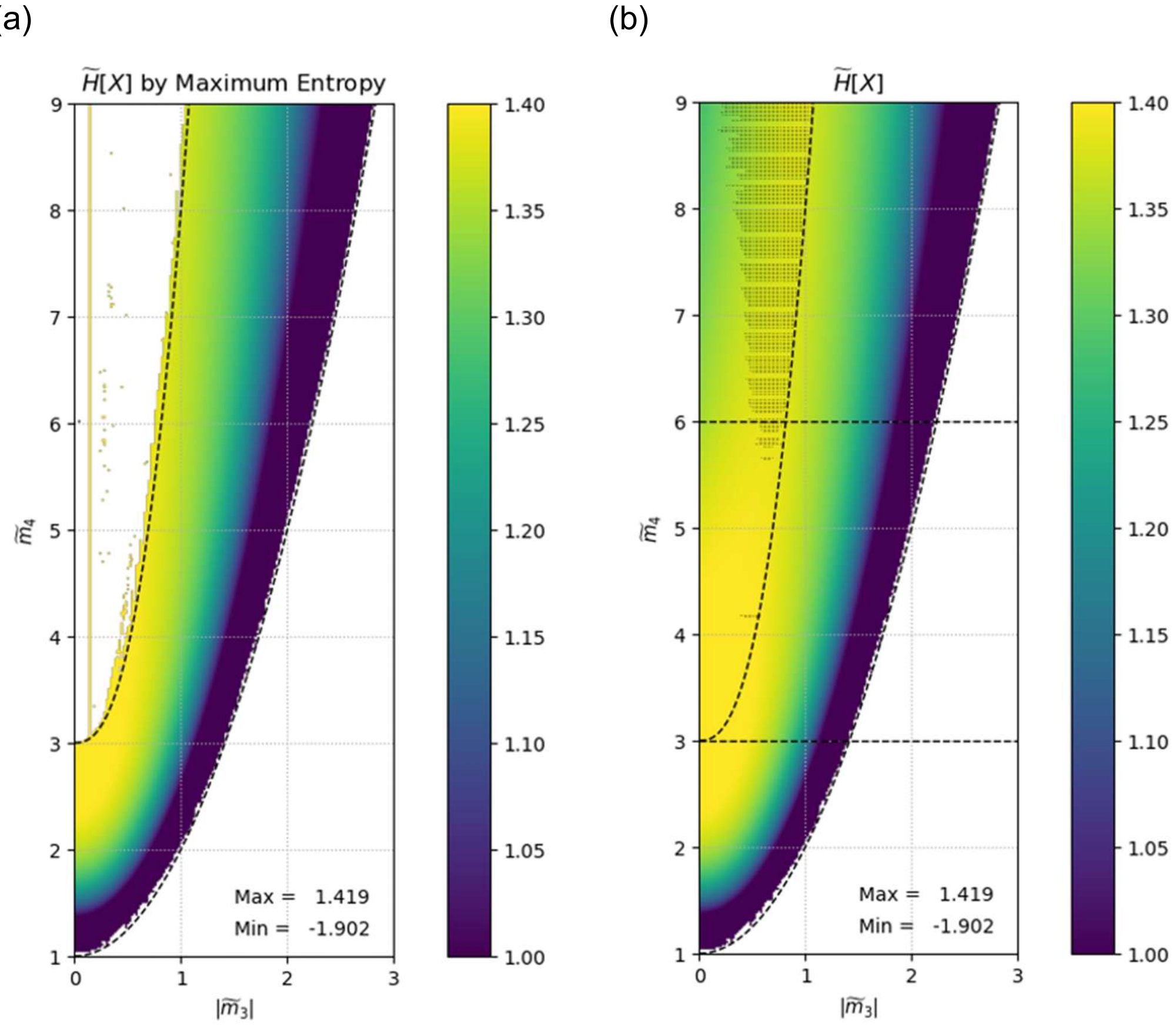

Fig. 2. Lookup tables for entropy: (a) a table obtained by using the maximum entropy distribution and (b) the table used in this study. In both panels, the upper dashed line indicates $\widetilde{m}_4 = 5\,|\widetilde{m}_3|^{2.5} + 3$, and the lower dashed line indicates $\widetilde{m}_4 = |\widetilde{m}_3|^2 + 1$. Dots in (b) represent the grid points where the derivative of $\widetilde{H}[\Delta\mathbf{z}^a(w)]$ with respect to $w$ is positive, and two horizontal dashed lines in (b) indicates $\widetilde{m}_4 = 3$ and 6.

Let us take a close look at the performance of this lookup table by taking an example shown in Table 1. The entropies estimated by table lookup are given in the "Table lookup" row along with the values of $\widetilde{m}_3^f$ and $\widetilde{m}_4^f$. These normalized moments are computed by using Eqs. (32) – (35), and $H[X|Y]$ is computed by using the identity of mutual information in Eq. (28). Since the maximum entropy method is adopted in the area between the two graphs $\widetilde{m}_4 = 5\,|\widetilde{m}_3|^{2.5} + 3$ and $\widetilde{m}_4 = |\widetilde{m}_3|^2 + 1$, those entropy values tend to be overestimations. If a distribution is close to be Gaussian such as $H_{\text{PO-E}}\ [X|Y]$, overestimation is negligible. Large overestimations occur in $H[X]$, $H[X|Y]$, and $H_{\text{ETKF}}[X|Y]$; the overestimation of $H[X|Y]$ is mainly caused by the overestimation of $H[X]$. Note that the magnitude of overestimation of $H_{\text{ETKF}}[X|Y]$ is the same as that of $H[X]$. Consequently, when we compare $H_{\text{ETKF}}[X|Y]$ with $H[X|Y]$, those two overestimations are cancelled. This observation justifies the choice of LETKF as the first guess in solving Eq. (44). However, as the iterations of Newton's method progress, the overestimation of the left-hand side of Eq. (44) is reduced due to increasing Gaussianity, whereas the overestimation on the right-hand side remains unchanged. As a result, the solution to Eq. (44) may be underestimated. This is a problem introduced by the lookup table based on maximum entropy distributions, but the overestimation of $H[X|Y]$ in Table 1 is only 3 % and it may not be a serious problem.

However, there is a severe problem with the use of third- and fourth-order moments estimated from a finite ensemble. The empirical estimation of these moments is notoriously robust-deficient, as a few extreme members can disproportionately distort the results. Let us take the posterior PDF in Fig. 1 as an example. This PDF seems close to a Gaussian distribution, but it has the secondary peak of 0.002 at $x = -2.6$, and the normalized third- and fourth-order central moments are –1.3 and 11.0, respectively. These values are quite different from those of Gaussian distribution, which are 0 and 3, respectively. Consequently, the third- and fourth-order moments estimated from a finite ensemble following this posterior PDF drastically vary depending on whether an extreme member emerges around the secondary peak or not. Those moments also greatly vary depending on the location of the extreme member. In this study, we do not place much confidence on large values of $\tilde{m}_4^f$. Specifically, we introduce a threshold value $\tilde{m}_4^c$ for $\tilde{m}_4^f$; if $\tilde{m}_4^c < \tilde{m}_4^f < \tilde{m}_4^c + 3$ then we take linear interpolation of the optimized weight parameter and the weight parameter of LPO-EnKF, and if $\tilde{m}_4^f \geq \tilde{m}_4^c + 3$ then we adopt the weight parameter of LPO-EnKF. The use of LPO-EnKF for large values of $\tilde{m}_4^f$ may not be the best choice, but it is a better choice than the LETKF because the LPO-EnKF is more robust to non-Gaussianity.

## 4. Data assimilation experiments

We conduct data assimilation experiments using the 40-variable Lorenz-96 model to confirm the validity of the optimization method of MI-EnKF for a case where the observation operator is linear (linear case) and a case where it is strongly nonlinear (nonlinear case).

### *4.1. Experimental design*

a. Model

The governing equation of the Lorenz-96 model is

$$\frac{dx_k}{dt} = -x_{k-1}(x_{k-2} - x_{k+1}) - x_k + F \tag{46}$$

for $k = 1, \cdots, K$, satisfying periodic boundary conditions: $x_{-1} = x_{K-1},\ x_0 = x_K,\ x_{K+1} = x_1$ The number of state variables $K$ and the forcing parameter $F$ are set to the conventional values: $K = 40$ and $F = 8$. According to Lorenz and Emanuel (1998), the number of positive Lyapunov exponents of the model is 13, and the fractional dimension of the attractor, as estimated from the formula of Kaplan and Yorke (1979), is about 27.1. The leading Lyapunov exponent corresponds to a doubling time of 0.42.

Time integration of the model provides true states that are used for the verification of analysis. The fourth-order Runge-Kutta scheme is used for time integration with a time step 0.01. The initial condition at each grid point is set to $F$ plus an independent random number drawn from a Gaussian distribution with mean 0 and variance 4. The model is integrated from $t =$ 0 to $t =$ 1 050.

b. Observations

In the linear case, the state variable is directly observed, that is, $\mathcal{H}(\boldsymbol{x}) = \boldsymbol{x}$. In the nonlinear case, we adopt a strongly nonlinear observation operator examined by Poterjoy (2006): $\mathcal{H}(\boldsymbol{x}) = \log |\boldsymbol{x}|$, where absolute value and logarithmic operations are performed on each element of $\boldsymbol{x}$. For both cases, the observational data are generated by adding Gaussian random errors with mean 0 and variance 1 to true states, and they are provided at each grid point with a time interval 0.05. As this time interval is much smaller than the time scale of the leading Lyapunov exponent, data assimilation is conducted at a high frequency and the time evolution of the model between the adjacent assimilation times is almost linear.

c. Data assimilation settings

We conduct data assimilation experiments using the LETKF, the LPO-EnKF, and the MI-EnKF with $d_c = 1$, 2, and 3. The threshold $\widetilde{m}_4^c$ is set to 3 and 6; these values are shown in Fig. 2b by horizontal dashed lines. The observation error covariance matrix $\boldsymbol{R}$ is set to $\boldsymbol{I}_{40}$. Data assimilation is started at $t =$ 0, and the analyses from $t$ = 0 through $t$ = 50 are not used for verification to avoid adverse effects of spin up. Analysis accuracy is estimated by the average root mean square error (RMSE). It is the square root of the square error averaged over the grid points and the verification period $50.05 \leq t \leq 1050$; the number of samples is 20 000. The ensemble size is set to 10 and 40 for the linear case, and 10, 20, 30, and 40 for the nonlinear case. The ensemble size of 10 is not very small compared with the number of positive Lyapunov exponents of the Lorenz-96 model, and an ensemble size of 40 is the same as the degrees of freedom of the model.

For LETKF (LPO-EnKF) runs, we set $w_i$ to 1 ($1/\sqrt{1+\sigma_i^2}$) in Eq. (19) for $i = 1, \cdots, d$. For MI-EnKF runs, we solve Eq. (44) when $\widetilde{m}_4^f < \widetilde{m}_4^c + 3$ for $i = 1, \cdots, d_c$, using Newton's method and the lookup table for entropy. The two derivatives on the right-hand side of Eq. (45) are estimated from this table by using a finite difference method. The first guess of $w$ is set to 1, and we solve Eq. (44) only if $\widetilde{H}[\Delta \boldsymbol{z}^a(1)]$ is smaller than the right-hand side of this equation. The convergence criterion is given as $|\Delta w|$ < 0.001, where $\Delta w$ is the correction term to $w$ in each iteration step of Newton's method. The maximum number of iterations is set to 10. Convergence occurred at a couple of iterations in most cases of the data assimilation experiments of this study. When $\tilde{m}_4^f > \tilde{m}_4^c$ for $i = 1, \cdots, d_c$, we employ the procedure mentioned at the end of Subsection 3.3. We set $w_i$ to 1 for $i = d_c + 1, \cdots, d$.

An EnKF needs covariance localization and covariance inflation to optimize its performance. The matrix $\boldsymbol{R}^{-1}$ that appears in Subsection 2.1 is localized according to the R-localization, in which the correlation function defined by Eq. (4.10) of Gaspari and Cohn (1999) is adopted. The parameter $c$ in this equation is regarded as the localization radius $r_L$ (unit: grid interval), at which radius the correlation coefficient decreases to 5/24. The radius of a local domain is also set to $r_L$. The value of $r_L$ is changed in increments of 1 from 1 to 19 in the linear case, and from 1 to 10 in the nonlinear case. The number of observations in a local domain is given by $2r_L + 1$. Note that $\{\boldsymbol{f}_i^o\}_{i=1}^d$ in Eq. (19) are not affected by the R-localization, because the covariance matrix of $\left\{\boldsymbol{\varepsilon}^{o(k)}\right\}_{k=1}^N$ in Eq. (15) is given by $(1 - 1/N)\boldsymbol{R}$.

The adaptive multiplicative inflation proposed by Li et al. (2009) is used to inflate the forecast ensemble when calculating the analysis ensemble in a local domain. This method is based on the innovation statistics by Desroziers et al. (2005). In the nonlinear case, we apply it to the observed variable $\mathcal{H}(\boldsymbol{x}) = \log|\boldsymbol{x}|$ and adopt the resulting inflation factor as the inflation factor for $\boldsymbol{x}$. This procedure can be regarded as employing a nonlinear extension of tangent linear approximation. Li et al. (2009) imposed lower and upper bounds in the "observed" inflation factor $\tilde{\rho}$ before applying a smoothing procedure; they set the range of $\tilde{\rho}$ to [0.9, 1.2] in their data assimilation experiments using the Lorenz-96 model and the LETKF with a linear observation operator. Since we also conduct data assimilation experiments using the LPO-EnKF, the MI-EnKF, and the nonlinear observation operator, we try several values of the upper bound of $\tilde{\rho}$; $\tilde{\rho}_{max}$ is set to 1.2, 1.3, 1.4, and 1.5 with the lower bound left at the original value 0.9.

*4.2. Results of the linear case*

Figure 3 compares the average analysis RMSEs of LETKF, LPO-EnKF and MI-EnKF with $\tilde{m}_4^c = 3$ for $N = 10$ and 40 in the linear case. They are plotted against the localization radius $r_L$, and each EnKF adopts the optimal value of $\tilde{\rho}_{max}$, where the optimal value is the value at which the RMSE is minimized. There is almost no difference between MI-EnKFs with $\tilde{m}_4^c =$ 3 and 6 except for $r_L \geq 14$ of the $N = 10$ case. This figure reveals that the LETKF is superior to the LPO-EnKF in terms of analysis accuracy, and that the MI-EnKF has the same accuracy as the LETKF regardless of the number of optimized modes.

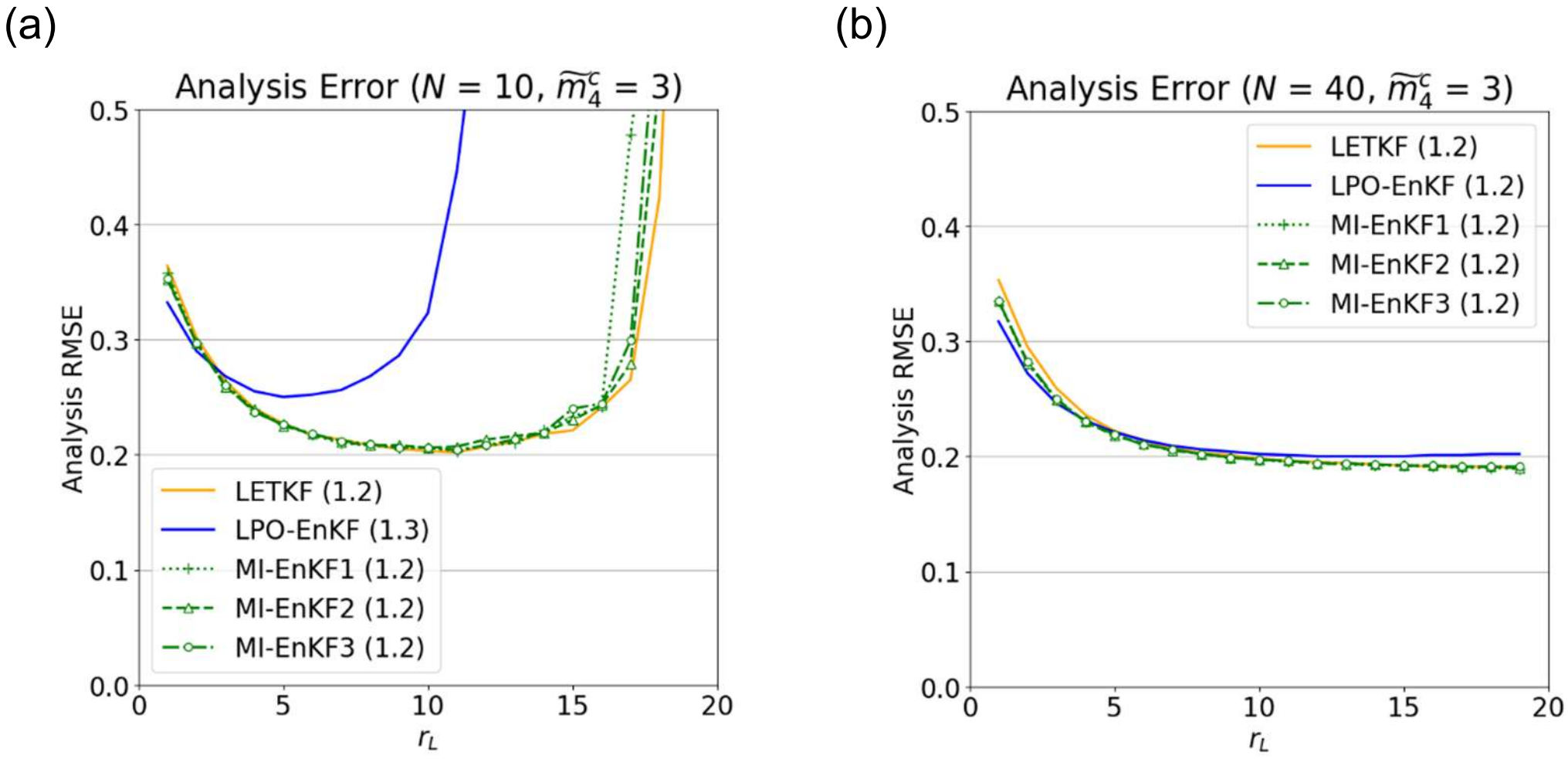

Fig. 3. Average analysis RMSEs plotted against $r_L$ of LETKF (orange line), LPO-EnKF (blue line), and MI-EnKF with $d_c = 1$ (dotted green line), $d_c = 2$ (dashed green line), and $d_c = 3$ (dash-dot green line) in the linear case for $\widetilde{m}_4^c = 3$: (a) $N = 10$ and (b) $N = 40$. The digits in parentheses in the legend are $\tilde{\rho}_{max}$, which are the optimal values.

Figure 4 displays the scatter diagrams of $|\widetilde{m}_3^f|$ and $\widetilde{m}_4^f$ of the first mode of MI-EnKF with $d_c = 3$, $\widetilde{m}_4^c = 3$, and the optimal values of $r_L$ and $\tilde{\rho}_{max}$. The sample points are extracted from the first grid point of the model domain at a time interval of 0.5, so that the number of sample points is 2 000. The two dashed graphs in each panel of this figure are the same as those in Fig. 2a, but the upper graph is plotted only in $\widetilde{m}_4^f \leq 9$ because it is an empirical graph for the domain of Fig. 2. Figure 4 reveals that approximately 10 % (30 %) of sample points exceed $\widetilde{m}_4^f = 3$ when $N = 10$ ($N = 40$) and that the sample points of $N = 40$ is slightly shifted towards larger $\widetilde{m}_4^f$ values compared to those of $N = 10$. This is because more extreme members tend to emerge with an increase of ensemble size and an extreme member tends to persist during data assimilation cycles at a high frequency, as discussed by Tsuyuki (2024). On the other hand, the spread in the direction of the $|\widetilde{m}_3^f|$ axis is decreased as the ensemble size is increased, indicating that the forecast ensemble tends to distribute more symmetrically. Those characteristics are also seen in the scatter diagrams of LETKF. It is expected that as the ensemble size approaches infinity the sample points concentrate at $(|\widetilde{m}_3^f|, \widetilde{m}_4^f) = (0, 3)$ and the forecast ensemble becomes almost always Gaussian.

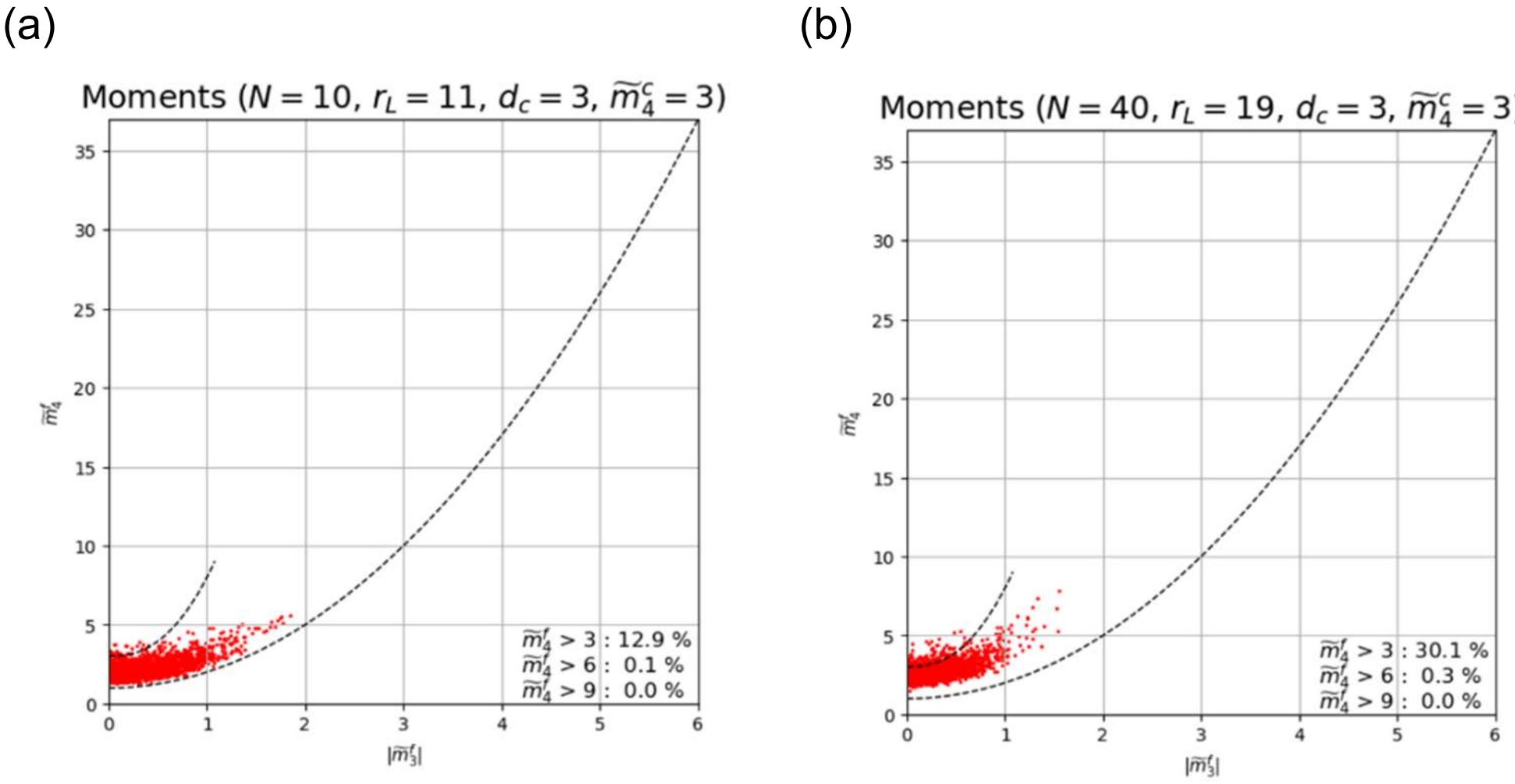

Fig. 4. Scatter diagrams between $|\widetilde{m}_3^f|$ and $\widetilde{m}_4^f$ of the first mode at the first grid point of MI-EnKF with $d_c = 3$ and the optimal values of $r_L$ in the linear case for $\widetilde{m}_4^c = 3$ and $\tilde{\rho}_{max} = 1.2$: (a) $N = 10$ and (b) $N = 40$. Dashed lines indicate the graphs of $\widetilde{m}_4 = 5\,|\widetilde{m}_3|^{2.5} + 3$ and $\widetilde{m}_4 = |\widetilde{m}_3|^2 + 1$.

The average eigenvalue spectra $\{\overline{\sigma_i^2}\}_{i=1}^{d}$ of MI-EnKF with the same parameter values as in Fig. 4 are presented in Fig. 5. The number of modes $d$ is given by $\min(N-1,\ 2r_L+1)$. This figure reveals that eigenvalues decrease exponentially with the mode number. This result justifies the procedure mentioned at the beginning of Subsection 3.2, because the weight parameter of LPO-EnKF, which is given by $1/\sqrt{1+\sigma_i^2}$, approaches that of LETKF as the mode number increases. The eigenvalues for $N = 10$ are smaller than those for $N = 40$ due to the difference in local domain size. The eigenvalue spectra for $\widetilde{m}_4^c = 6$ with the same values of $r_L$ are quite similar to Fig. 5.

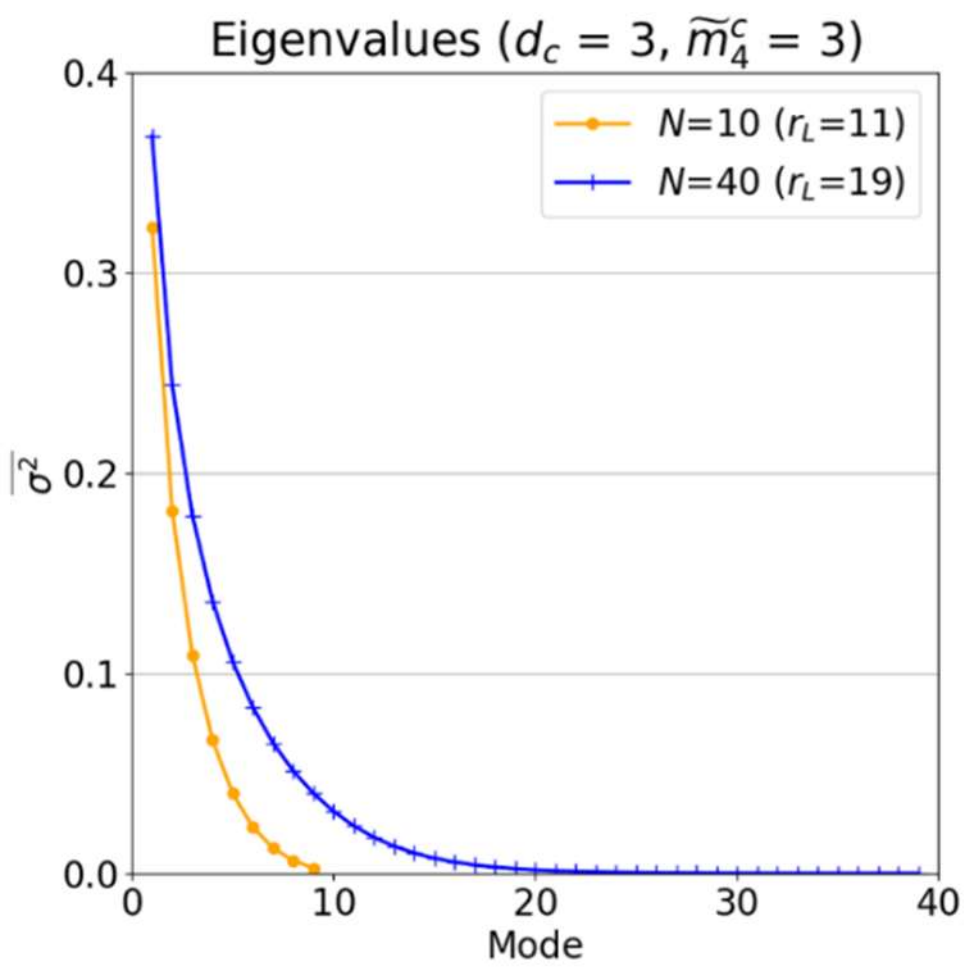

Fig. 5. Average eigenvalues of MI-EnKF with $d_c = 3$ and the optimal values of $r_L$ for $N = 10$ (orange line) and $N = 40$ (cyan line) in the linear case for $\widetilde{m}_4^c = 3$ and $\tilde{\rho}_{max} = 1.2$. They are plotted against the mode number.

Figure 6 shows the average weight parameters of MI-EnKF for $\widetilde{m}_4^c = 3$ and 6, along with those of LETKF and LPO-EnKF. Note that the plotted values of LPO-EnKF are not taken from LPO-EnKF runs but extracted from MI-EnKF runs. This figure reveals that the weight parameters of the MI-EnKF are close to 1, indicating that the MI-EnKF works like the LETKF. It also reveals that (i) the average weight parameters get closer to 1 with an increase of the mode number, (ii) those for $\widetilde{m}_4^c = 3$ are smaller than those for $\widetilde{m}_4^c = 6$, and (iii) those for $N = 10$ are larger than those for $N = 40$. The first and third results can be explained by the identity of mutual information in Eq. (28). As mentioned above, eigenvalues decrease exponentially with the mode number and eigenvalues for $N = 10$ are smaller than those for $N = 40$. Equation (41) indicates that a smaller eigenvalue enhances the difference between $(\langle\widetilde{m}_3^s\rangle, \widetilde{m}_4^s)$ and $(\langle\widetilde{m}_3^f\rangle, \widetilde{m}_4^f)$, which results in an increase of $\widetilde{H}[\Delta\boldsymbol{z}^f + \boldsymbol{f}^o] - \widetilde{H}[\Delta\boldsymbol{z}^f]$ due to more Gaussianity of $\Delta\boldsymbol{z}^f + \boldsymbol{f}^o$ and, therefore, results in a decrease of the difference between $\widetilde{H}[\Delta\boldsymbol{z}^a(1)]$ and the right-hand side of Eq. (44). Therefore, the weight parameter is closer to 1 with a decrease of eigenvalue. As for the second result, if $\widetilde{m}_4^c < \widetilde{m}_4 \leq \widetilde{m}_4^c + 3$, we take linear interpolation of the optimized weight parameter and the weight parameter of LPO-EnKF. Consequently, the weight parameters of $\widetilde{m}_4^c = 3$ are more influenced by the LPO-EnKF than those of $\widetilde{m}_4^c = 6$.

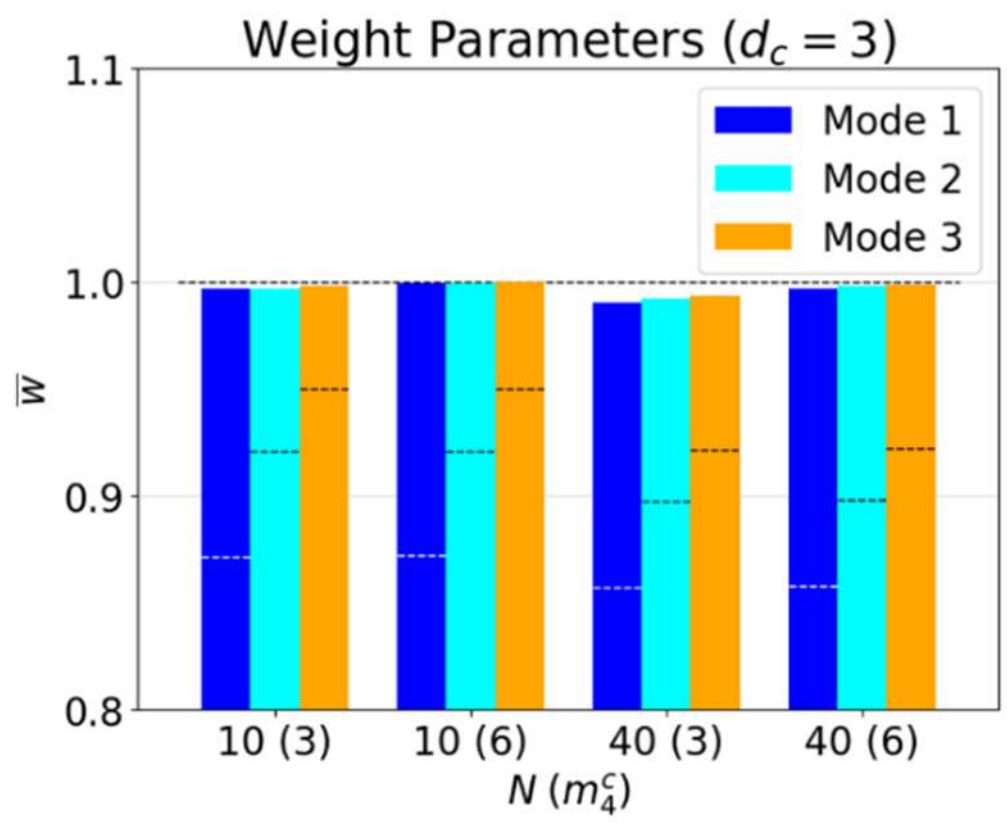

Fig. 6. Average weight parameters of MI-EnKF for the first three modes with $d_c = 3$ and the optimal values of $r_L$ in the linear case for $\tilde{\rho}_{max} = 1.2$. They are plotted against a combination of $N$ and $\tilde{m}_4^c$. Upper and lower dashed lines for each bar indicate the weight parameters of LETKF and the average weight parameters of LPO-EnKF extracted from MI-EnKF runs, respectively.

### *4.3. Results of the nonlinear case*

In this case, analysis accuracy intermittently becomes poor during data assimilation cycles due to the strong nonlinearity of observation operator and, therefore, results are more affected by sampling noise. The average analysis RMSEs of the LETKF, LPO-EnKF, and MI-EnKF with $d_c = 1$, 2, and 3 in the nonlinear case are plotted against the localization radius $r_L$ in Fig. 7 for $\tilde{m}_4^c = 3$ and in Fig. 8 for $\tilde{m}_4^c = 6$. When filter divergence occurs during data assimilation cycles, the RMSE is not plotted. Unlike Fig. 3, the adopted values of $\tilde{\rho}_{max}$ are representative ones that are set to the same value in each panel. Those figures reveal that the LPO-EnKF is superior to the LETKF in terms of analysis accuracy except for $N = 10$, but its accuracy is sensitive to the localization radius compared to the LETKF. As for the MI-EnKF, optimizing just the first mode can lead to significant improvements compared to the LETKF, and the MI-EnKF with $d_c = 3$ and $\tilde{m}_4^c = 3$ generally outperforms both LETKF and LPO-EnKF regardless of ensemble size. An exception is seen in Fig. 8b, which indicates that the MI-EnKF with $d_c = 2$ and $\tilde{m}_4^c = 6$ shows the highest accuracy. However, the MI-EnKF with $\tilde{m}_4^c = 6$ is generally less accurate, suggesting that the optimized weight parameters for large values of $\tilde{m}_4^f$ is less reliable. RMSE time series (not shown) reveal that the improvement in average accuracy is primarily due to significant reduction in instances of major accuracy degradation. Another advantage of MI-EnKF is that analysis accuracy is not sensitive to the localization radius, like the LETKF. An unexpected result is that positive impacts of increasing the number of optimized modes is not observed except for $N = 40$. Incidentally, in the case of $N = 10$, the accuracy of MI-EnKF is worse than that of LETKF when $r_L \geq 5$ for $\tilde{m}_4^c = 3$ and $r_L \geq 6$ for $\tilde{m}_4^c = 6$.

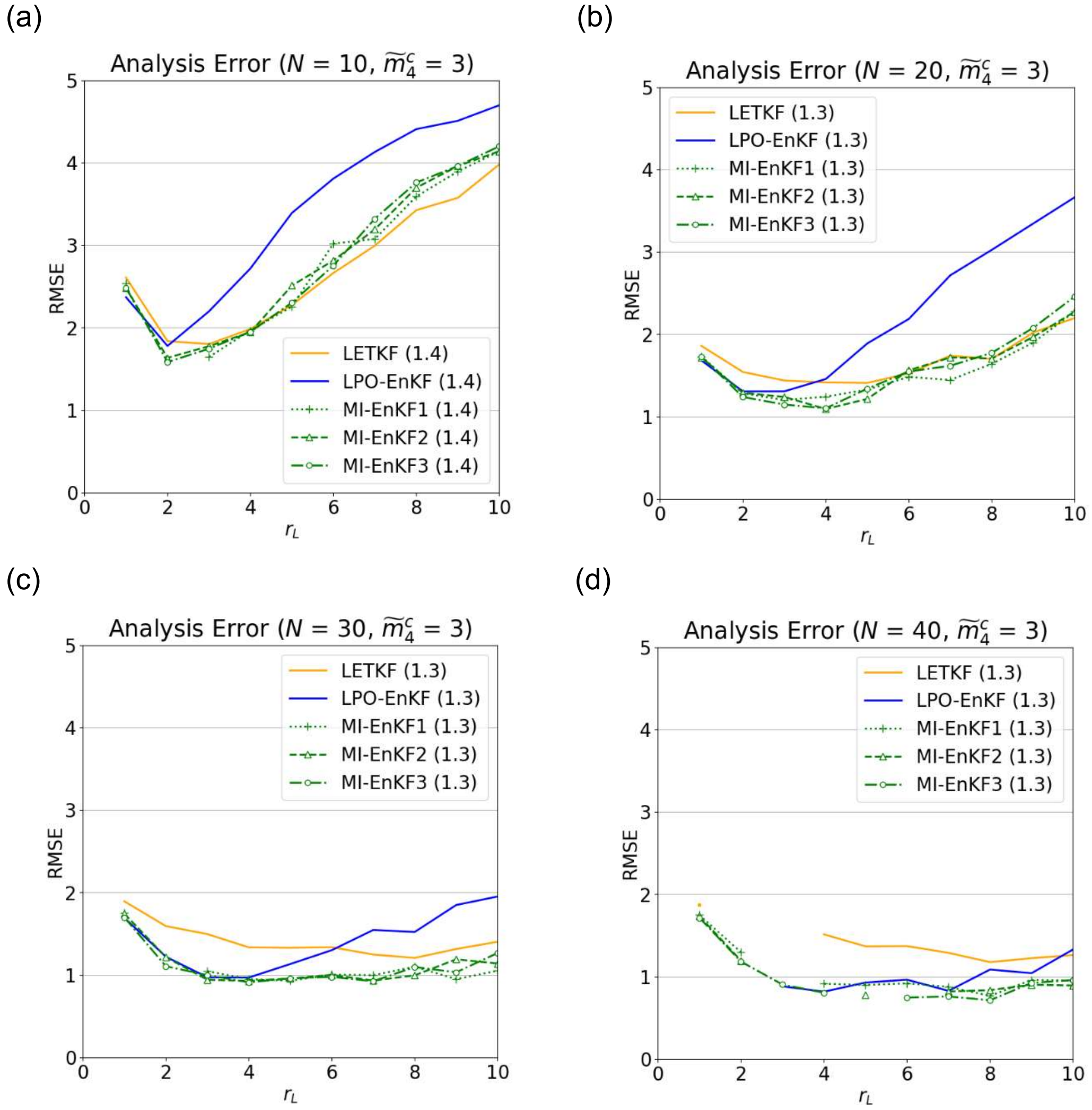

Fig. 7. Average analysis RMSEs plotted against $r_L$ of LETKF (orange line), LPO-EnKF (blue line), and MI-EnKF with $d_c = 1$ (dotted green line), $d_c = 2$ (dashed green line), and $d_c = 3$ (dash-dot green line) in the nonlinear case for $\widetilde{m}_4^c = 3$: (a) $N = 10$, (b) $N = 20$, (c) $N = 30$, and (d) $N = 40$. The digits in parentheses in the legend are $\tilde{\rho}_{max}$.

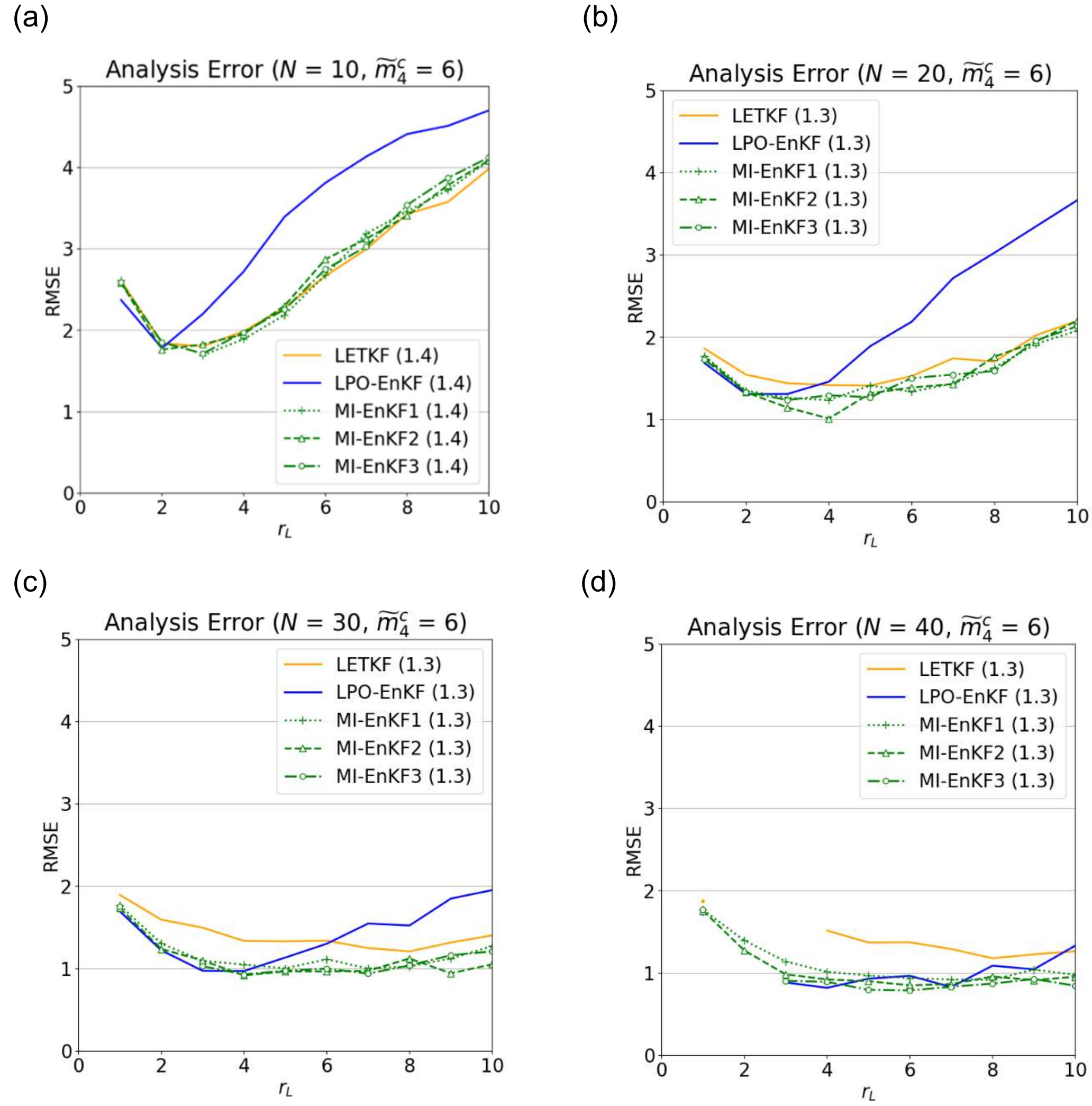

Fig. 8. Same as Fig. 7 except for $\widetilde{m}_4^c = 6$.

Figure 9 shows the average analysis RMSEs of the LETKF and LPO-EnKF, and the RMSE reduction by the MI-EnKF with $d_c = 3$ and $\widetilde{m}_4^c = 3$, which are plotted against $r_L$ and $\tilde{\rho}_{max}$. A grid box where filter divergence occurred is shown in gray. The RMSE reduction is defined as the difference between the smaller RMSE of LETKF and LPO-EnKF and the RMSE of MI-EnKF, so that a positive RMSE reduction means that the MI-EnKF is more accurate than both LETKF and LPO-EnKF. If either LETKF or LPO-EnKF diverged, the RMSE reduction was calculated using one of the two EnKFs that did not diverge. Accordingly, a gray grid box in the right column of Fig. 9 indicates either that the MI-EnKF diverged or that both LETKF and LPO-EnKF diverged. The minimum RMSE in each panel is shown by red digits, and the minimum RMSE for each value of $\tilde{\rho}_{max}$ is shown by black digits. Note that the digits in the panels of RMSE reduction (right column of Fig. 9) indicate the RMSEs of MI-EnKF, not the RMSE reduction. This figure reveals again that the LPO-EnKF is superior to the LETKF except for $N = 10$ and that the MI-EnKF is the most accurate regardless of ensemble size. The negative RMSE reduction is widely seen when $N = 10$ (Fig. 9c), but this area decreases as the ensemble size increases. This is probably because the area where the analysis accuracy of MI-EnKF deteriorates gradually shifts towards larger $r_L$ values with an increase of ensemble size.

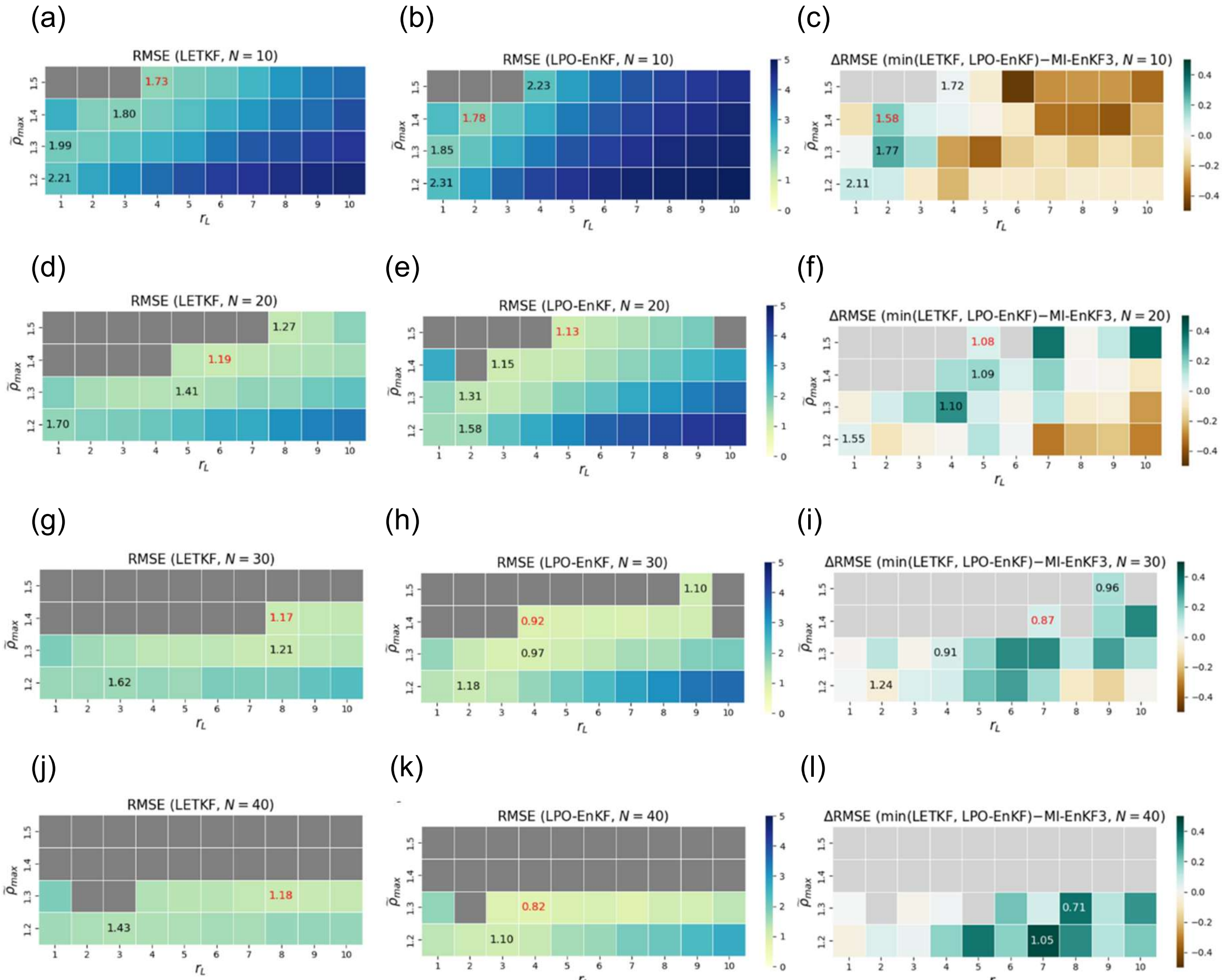

Fig. 9. Average analysis RMSEs of LETKF (left column) and LPO-EnKF (middle column), and average RMSE reduction of MI-EnKF with $d_c = 3$ and $\widetilde{m}_4^c = 3$ (right column) for $N = 10$, 20, 30, and 40 (from top to bottom) in the nonlinear case. The RMSE reduction is defined as the difference between the smaller of RMSEs of LETKF and LPO-EnKF and the RMSE of MI-EnKF. They are plotted against $r_L$ and $\tilde{\rho}_{max}$, and a grid box where filter divergence occurred is shown in gray. In each panel, the minimum RMSE in each panel is shown by red digits, and the minimum RMSE for each value of $\tilde{\rho}_{max}$ is shown by black digits, except for panel (l) where they are indicated by white digits to improve visibility.

The scatter diagrams of $|\tilde{m}_3^f|$ and $\tilde{m}_4^f$ of the first mode of MI-EnKF with the optimal values of $r_L$ are shown in Fig. 10 in a similar manner to Fig. 4. The parameter values are the same as in Fig. 7. A significant difference from the linear case is that the sample points are more widely distributed along the graph of $\tilde{m}_4 = |\tilde{m}_3|^2 + 1$ as the ensemble size is increased, implying that extreme members tend to distribute more asymmetrically in a histogram of the forecast ensemble. Another significant difference is that more extreme members emerge; approximately 40 % (80 %) of sample points exceed $\tilde{m}_4^f = 3$ when $N = 10$ ($N = 30$ and 40). As was observed in the linear case, the sample points is shifted towards larger $\tilde{m}_4^f$ values with an increase of ensemble size. Those characteristics are also seen in the scatter diagrams of LETKF.

The average eigenvalue spectra $\{\overline{\sigma_i^2}\}_{i=1}^{d}$ of MI-EnKF are presented in Fig. 11 for both $\tilde{m}_4^c = 3$ and $\tilde{m}_4^c = 6$. The parameter values are same as in Fig. 10. The eigenvalues are much larger than those in the linear case (Fig. 5). Similar characteristics to those observed in the linear case are observed except that when $\tilde{m}_4^c = 6$ the eigenvalue of the first mode for $N = 10$ is very close to that for $N = 40$. This may be caused by sampling noise.

Figure 12 shows the average weight parameters of MI-EnKF for $\tilde{m}_4^c = 3$ and 6, along with those of LETKF and LPO-EnKF. The parameter values are the same as in Fig. 10. This figure reveals that the optimal EnKF lies between the LETKF and the LPO-EnKF. The weight parameters in the nonlinear case are smaller than those in the linear case (Fig. 6), indicating that stochastic term has more weight in the analysis ensemble. It also reveals that the results (i), (ii), and (iii) that was mentioned in the last paragraph of Subsection 4.2 also hold in the nonlinear case. The first mode has the largest eigenvalue and the smallest weight parameter. This result explains why optimizing just the first mode can lead to significant improvements.

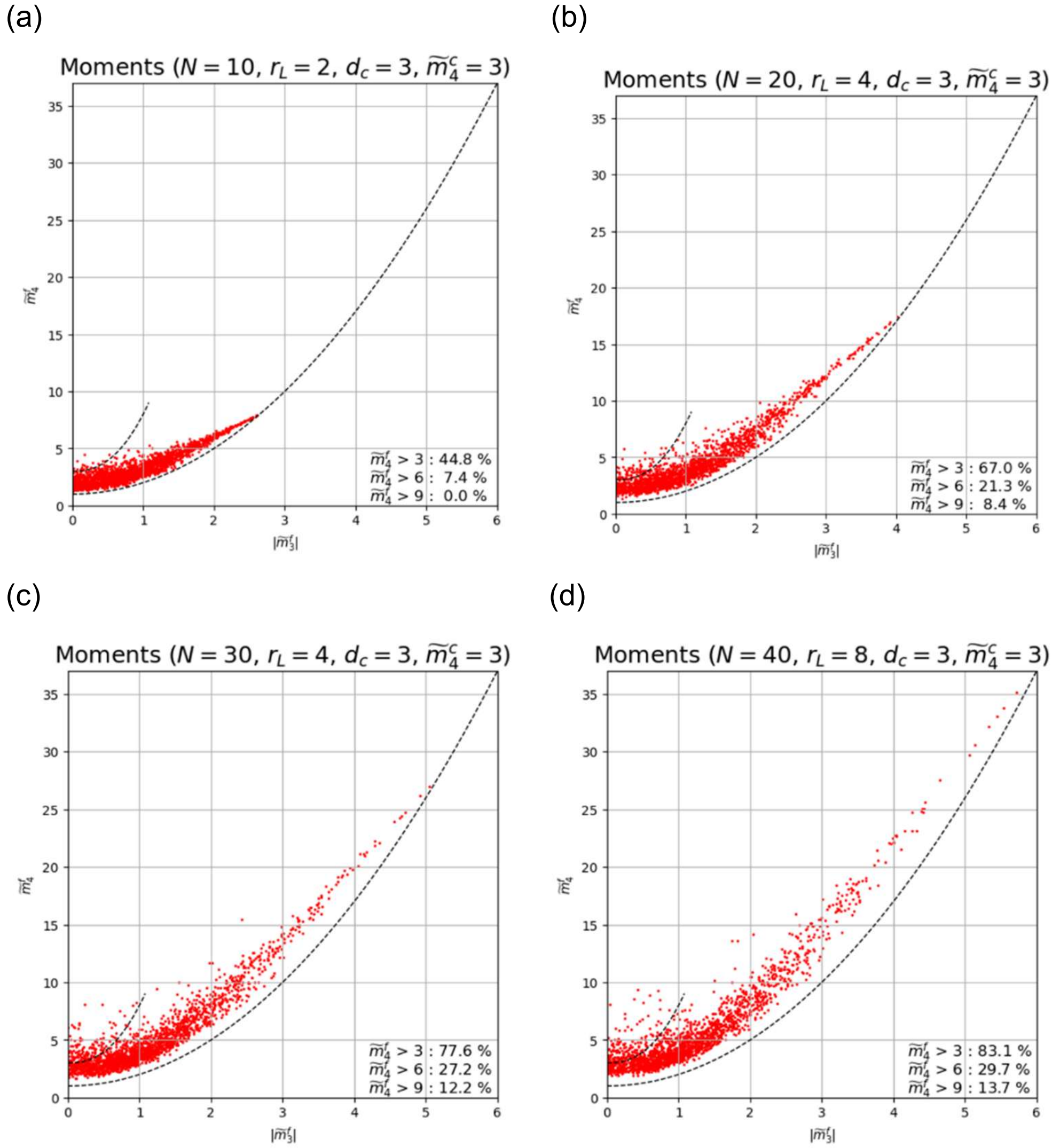

Fig. 10. Scatter diagrams between $\left|\widetilde{m}_3^f\right|$ and $\widetilde{m}_4^f$ of the first mode at the first grid point of MI-EnKF with $d_c = 3$, $\widetilde{m}_4^c = 3$ and the optimal values of $r_L$ in the nonlinear case: (a) $N = 10$ and $\tilde{\rho}_{max} = 1.4$, (b) $N = 20$ and $\tilde{\rho}_{max} = 1.3$, (c) $N = 30$ and $\tilde{\rho}_{max} = 1.3$, and (d) $N = 40$ and $\tilde{\rho}_{max} = 1.3$. Dashed lines indicate the graphs of $\widetilde{m}_4 = 5\,|\widetilde{m}_3|^{2.5} + 3$ and $\widetilde{m}_4 = |\widetilde{m}_3|^2 + 1$.

(a) (b)

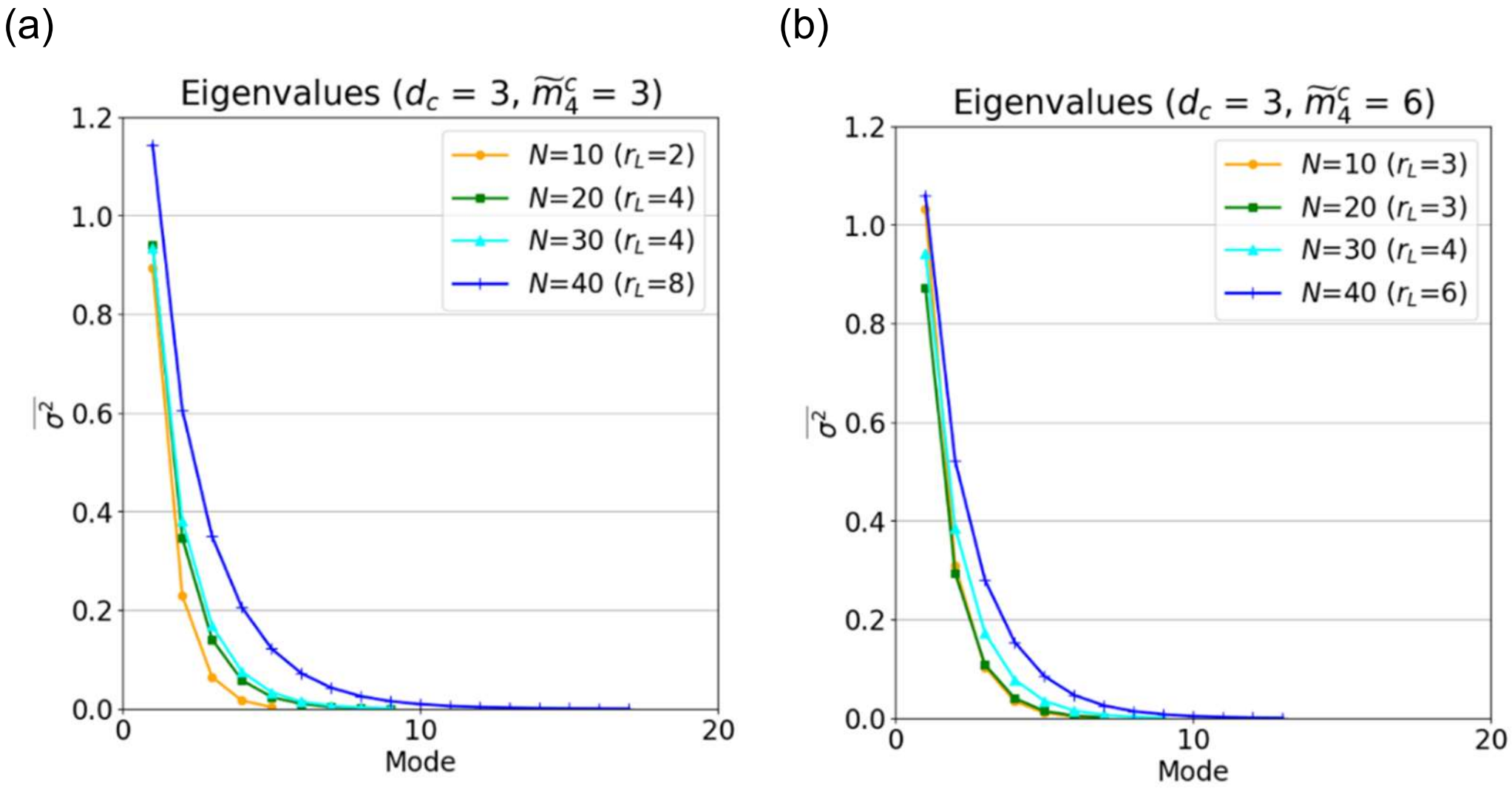

Fig. 11. Average eigenvalues of MI-EnKF with $d_c = 3$ and the optimal values of $r_L$ for $N = 10$ and $\tilde{\rho}_{max} = 1.4$ (orange line), $N = 20$ and $\tilde{\rho}_{max} = 1.3$ (blue line), $N = 30$ and $\tilde{\rho}_{max} = 1.3$ (green line), and $N = 40$ and $\tilde{\rho}_{max} = 1.3$ (cyan line) in the nonlinear case: (a) $\widetilde{m}_4^c = 3$ and (b) $\widetilde{m}_4^c = 6$.

(a) (b)

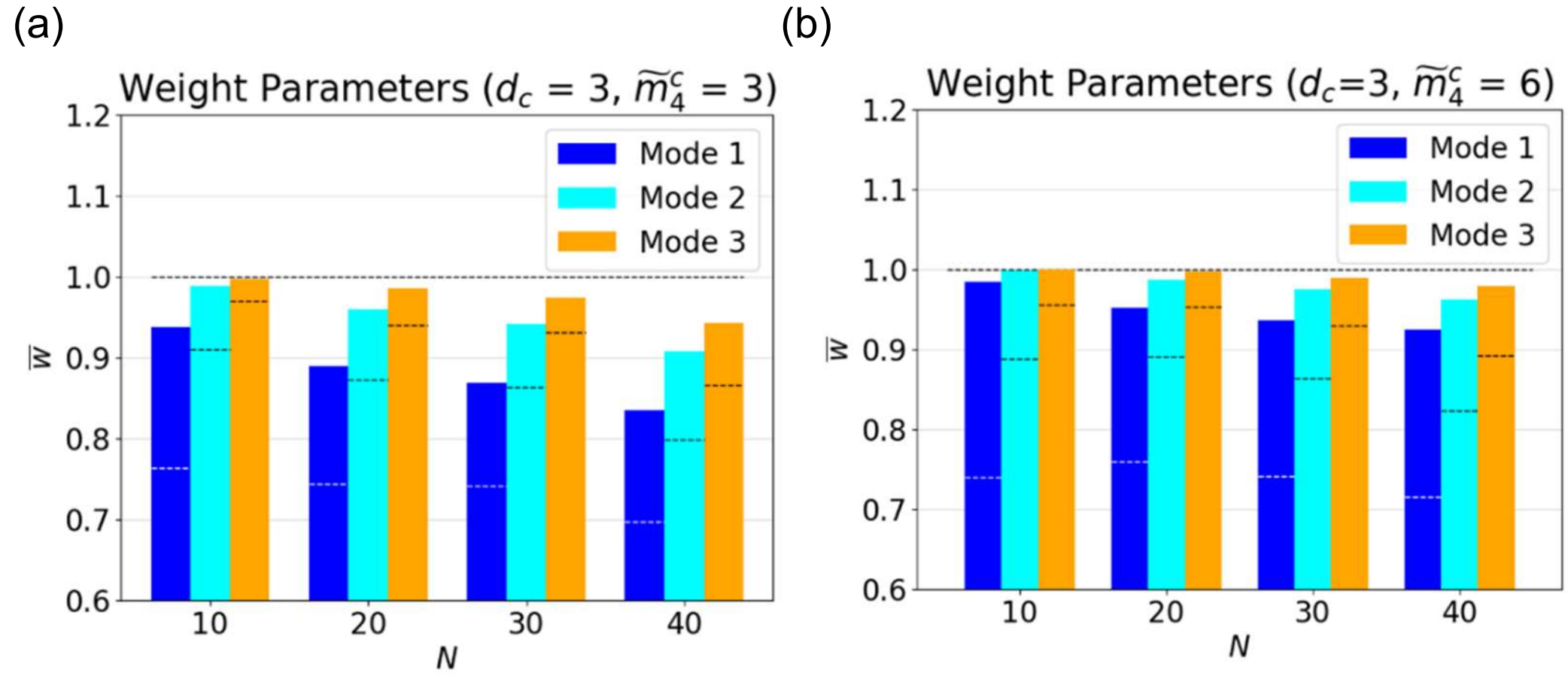

Fig. 12. Average weight parameters of MI-EnKF with $d_c = 3$ and the optimal values of $r_L$ for the first three modes in the nonlinear case: (a) $\widetilde{m}_4^c = 3$ and (b) $\widetilde{m}_4^c = 6$. The parameter values are the same as in Fig, 11. Upper and lower dashed lines for each bar indicate the weight parameters of LETKF and the average weight parameters of LPO-EnKF extracted from MI-EnKF runs, respectively.

As mentioned at the end of Subsection 3.3, when $\widetilde{m}_4^f > \widetilde{m}_4^c$ we used a weighting average of the optimized weight parameter and the weight parameter of LPO-EnKF or the latter parameter itself as the weight parameter of MI-EnKF. In other words, we treated the LPO-EnKF as the default data assimilation method for large values of $\widetilde{m}_4^f$. However, it might be better to adopt the MI-EnKF with the average weight parameters shown in Fig. 12 as the default data assimilation method for large values of $\widetilde{m}_4^f$. So, we conduct additional data assimilation experiments to compare the above two default settings for $\widetilde{m}_4^c = 3$ using another set of true states and observations. Figure 13 shows the average analysis RMSEs of the LETKF, LPO-EnKF, and MI-EnKFs using the two default settings with $d_c = 1$, 2, and 3. Features similar to those in Fig. 7 can be observed, although the MI-EnKF for $N = 40$ performs more stable. Although there is not much difference in analysis accuracy between the LPO-EnKF weight parameters and the average weight parameters, when $N = 40$ the MI-EnKF with the former parameters is consistently more accurate than that with the latter. This result confirms the robustness of stochastic EnKF in strongly nonlinear regimes of data assimilation.

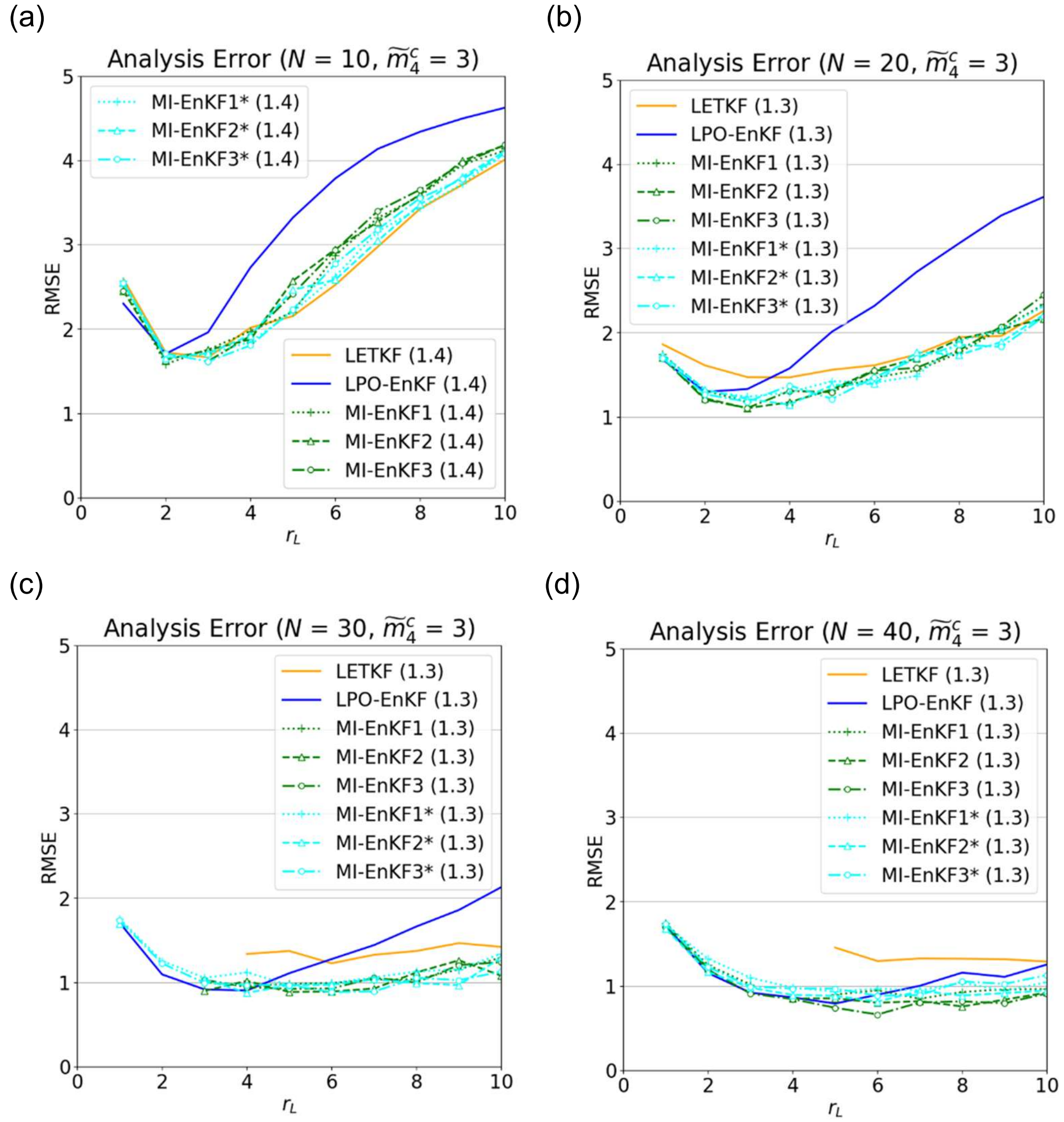

Fig. 13. Same as Fig. 7 except for addition of RMSEs of MI-EnKF using the average weight parameters for large values of $\widetilde{m}_4^f$ with $d_c = 1$ (dotted cyan line), $d_c = 2$ (dashed cyan line), and $d_c = 3$ (dash-dot cyan line) and for different true states and observations.

## 5. Summary and discussion

Tsuyuki (2024) showed that the analysis perturbation equations of EnKFs are decomposed into a set of equations for one-dimensional systems that have no correlations with each other. Based on this result, we generalize the LETKF such that it contains a stochastic term and includes both LETKF and LPO-EnKF within it. A generalized LETKF thus obtained has a parameter that determines the relative weight of the stochastic term. We adaptively optimize this parameter based on an identity of mutual information, which is satisfied by the Kalman filter in linear Gaussian systems. As the analysis perturbation equations of EnKFs are decomposed into a system of equations for modes that are uncorrelated with each other, the application of mutual information is easily achieved. The generalized LETKF thus optimized is named the MI-EnKF, which indirectly uses the third- and fourth-order moments of the forecast ensemble through entropy. To speed up calculations of entropy, we create a lookup table based on maximum entropy distributions.

We conduct data assimilation experiments using the Lorenz-96 model to confirm the validity of the optimization method of MI-EnKF for the linear case where the observation operator is linear and the nonlinear case where it is strongly nonlinear. The weight parameters of up to the first three modes are optimized. In the linear case, it is found that the analysis accuracy of MI-EnKF is the same as that of LETKF, which is more accurate than the LPO-EnKF. In the nonlinear case, the MI-EnKF is more accurate than both LETKF and LPO-EnKF regardless of ensemble size, and its analysis accuracy is less sensitive to the localization radius than that of LPO-EnKF, which is more accurate than the LETKF. Optimizing just the first mode can lead to significant improvements, but positive impacts of increasing the number of optimized modes are not observed unless the ensemble size is large. The optimized parameter values indicate that the optimal EnKF lies between the deterministic EnKF and the stochastic EnKF. Those results confirm the validity of the optimization method used in the MI-EnKF and demonstrate the effectiveness of information theory in advancing data assimilation methods.

The quality of a lookup table for entropy is important for the MI-EnKF to perform as expected. In this study, when Newton's method failed to converge, entropy was computed by using mixture distributions generated by linear interpolation of maximum entropy distributions, but questionable values were obtained in the dotted areas in Fig. 2b. Adopting a more accurate interpolation method may help to alleviate this problem, but a more promising approach would be to leverage generative artificial intelligence (AI). We could obtain a surrogate of maximum entropy from the samples of smooth distributions generated by a diffusion model (Song and Ermon 2019, Ho et al. 2020) with a constraint of the first four moments.

In the data assimilation experiments of this study, we introduced nonlinearity through the observation operator, since data assimilation was conducted at a high frequency. As the optimization method of MI-EnKF is based on the moments of the forecast ensemble, the MI-EnKF can also handle the nonlinearity of numerical models. Most of the researchers studying the LPF have used the LETKF as a comparison for the LPF. The MI-EnKF can be used as a fair comparison with the LPF due to its higher accuracy than the LETKF in strongly nonlinear regimes and its high computational efficiency like the LETKF.

**Acknowledgements**

This study was supported by the Japan Science and Technology Agency Moonshot R&D (JPMJMS2389), the Japan Society for the Promotion of Science (JSPS) via KAKENHI (grant no. JP24H00278 and JP25H00752), the IAAR Research Support Program of Chiba University, and the Joint Support-Center for Data Science Research via ROIS-DS-Joint (024RP2024 and 038RP2025) to T. Kawabata of the Meteorological Research Institute.

## Appendix 1. Kalman filtering in a mutual information perspective

The Kalman gain for linear Gaussian systems is usually derived by minimizing the trace of the analysis error covariance matrix, but it can also be derived by maximizing the mutual information between the state variable and the observation. In a linear Gaussian system, we can assume that the analysis vector $\boldsymbol{x}^a$ is a linear function of the forecast vector $\boldsymbol{x}^f$ and the observation vector $\boldsymbol{y}^o$:

$$\boldsymbol{x}^a = \boldsymbol{x}^f + \boldsymbol{W}\left(\boldsymbol{y}^o - \boldsymbol{H}\boldsymbol{x}^f\right), \tag{A1}$$

where $\boldsymbol{W}$ is a weight matrix and $\boldsymbol{H}$ is the observation operator matrix. Let us introduce two random variables $X$ and $Y$ that correspond to the state variable and the observational data, respectively. The mutual information between them is computed as

$$I[X,Y] = H[X] - H[X|Y] = \frac{1}{2}\log\frac{\left|\boldsymbol{P}^f\right|}{\left|\boldsymbol{P}^a\right|}, \tag{A2}$$

where $\boldsymbol{P}^f$ and $\boldsymbol{P}^a$ are the forecast and analysis error covariance matrices, respectively. The latter matrix is given by

$$\boldsymbol{P}^a = \boldsymbol{P}^f + \boldsymbol{W}\left(\boldsymbol{R} + \boldsymbol{H}\boldsymbol{P}^f\boldsymbol{H}^{\mathrm{T}}\right)\boldsymbol{W}^{\mathrm{T}} - \boldsymbol{P}^f\boldsymbol{H}^{\mathrm{T}}\boldsymbol{W}^{\mathrm{T}} - \boldsymbol{W}\boldsymbol{H}\boldsymbol{P}^f, \tag{A3}$$

where $\boldsymbol{R}$ is the observation error covariance matrix. Substitution of Eq. (A3) into Eq. (A2) and differentiation with respect of $\boldsymbol{W}$ yields

$$\frac{\partial I[X,Y]}{\partial \boldsymbol{W}} = -\left(\boldsymbol{P}^a\right)^{-1}\left[\boldsymbol{W}\left(\boldsymbol{R} + \boldsymbol{H}\boldsymbol{P}^f\boldsymbol{H}^{\mathrm{T}}\right) - \boldsymbol{P}^f\boldsymbol{H}^{\mathrm{T}}\right]. \tag{A4}$$

We obtain the following optimal weight matrix:

$$\boldsymbol{W}_{OPT} = \boldsymbol{P}^f\boldsymbol{H}^{\mathrm{T}}\left(\boldsymbol{R} + \boldsymbol{H}\boldsymbol{P}^f\boldsymbol{H}^{\mathrm{T}}\right)^{-1}. \tag{A5}$$

This is the Kalman gain.

Next, we show that the Kalman filter satisfies the identity of mutual information in Eq. (28) for linear Gaussian systems. Let the state variable be $n$-dimensional, and the observation be $m$-dimensional. Then, the mutual information of Kalman filter is calculated from Eq. (A2) as

$$H[X] - H[X|Y] = \frac{1}{2}\log\frac{\left|\boldsymbol{P}^f\right|}{\left|\left(\left(\boldsymbol{P}^f\right)^{-1} + \boldsymbol{H}^{\mathrm{T}}\boldsymbol{R}^{-1}\boldsymbol{H}\right)^{-1}\right|} = \frac{1}{2}\log\left|\boldsymbol{I}_n + \boldsymbol{H}^{\mathrm{T}}\boldsymbol{R}^{-1}\boldsymbol{H}\boldsymbol{P}^f\right|, \tag{A6}$$

where $\boldsymbol{I}_n$ is the $n$-dimensional identity matrix. On the other hand, the rightmost term of Eq. (28) is calculated as

$$H[Y] - H[Y|X] = \frac{1}{2}\log\frac{\left|\boldsymbol{R} + \boldsymbol{H}\boldsymbol{P}^f\boldsymbol{H}^{\mathrm{T}}\right|}{|\boldsymbol{R}|} = \frac{1}{2}\log\left|\boldsymbol{I}_m + \boldsymbol{H}\boldsymbol{P}^f\boldsymbol{H}^{\mathrm{T}}\boldsymbol{R}^{-1}\right|. \tag{A7}$$

Since $|\boldsymbol{I}_m + \boldsymbol{A}\boldsymbol{B}| = |\boldsymbol{I}_n + \boldsymbol{B}\boldsymbol{A}|$ for any $m \times n$ matrix $\boldsymbol{A}$ and $n \times m$ matrix $\boldsymbol{B}$, it is confirmed that the identity of mutual information in Eq. (28) holds. The above identity of determinants is easily derived from the following identity:

$$\begin{pmatrix} \boldsymbol{I}_m & \boldsymbol{A} \\ -\boldsymbol{B} & \boldsymbol{I}_n \end{pmatrix} = \begin{pmatrix} \boldsymbol{I}_m + \boldsymbol{AB} & \boldsymbol{A} \\ \boldsymbol{O}_{n\times m} & \boldsymbol{I}_n \end{pmatrix} \begin{pmatrix} \boldsymbol{I}_m & \boldsymbol{O}_{m\times n} \\ -\boldsymbol{B} & \boldsymbol{I}_n \end{pmatrix} = \begin{pmatrix} \boldsymbol{I}_m & \boldsymbol{O}_{m\times n} \\ -\boldsymbol{B} & \boldsymbol{I}_n \end{pmatrix} \begin{pmatrix} \boldsymbol{I}_m & \boldsymbol{A} \\ \boldsymbol{O}_{n\times m} & \boldsymbol{I}_n + \boldsymbol{BA} \end{pmatrix}, \quad \text{(A8)}$$

where $\boldsymbol{O}_{m\times n}$ denote the $m \times n$ zero matrix.

## Appendix 2. Lookup table for entropy

We briefly describe the method of creating the lookup tables in Fig. 2. To solve Eq. (48), we introduce the following functions of $\boldsymbol{\lambda} := (\lambda_1, \lambda_2, \lambda_3, \lambda_4)^{\mathrm{T}}$:

$$G_l(\boldsymbol{\lambda}) := \int_{-\infty}^{\infty} x^l \exp\left(-\sum_{k=1}^{4} \lambda_k x^k\right) dx \qquad (l = 0, 1, 2, \cdots), \quad \text{(A9)}$$

which satisfy

$$\frac{\partial G_l(\boldsymbol{\lambda})}{\partial \lambda_k} = -G_{k+l}(\boldsymbol{\lambda}) \qquad (k = 1, \cdots, 4) \quad \text{(A10)}$$

$$\sum_{k=1}^{4} k\, \lambda_k\, G_{k+l}(\boldsymbol{\lambda}) = (l+1) G_l(\boldsymbol{\lambda}). \quad \text{(A11)}$$

The latter equation is derived by applying integration by parts to Eq. (A9). Because $\lambda_0 = \log G_0(\boldsymbol{\lambda})$, Eq. (48) is rewritten as

$$G_l(\boldsymbol{\lambda}) = m_l G_0(\boldsymbol{\lambda}) \qquad (l = 1, \cdots, 4). \quad \text{(A12)}$$

This set of equations is numerically solved using Newton's method. The correction terms $\{\Delta\lambda_k\}_{k=1}^4$ to $\boldsymbol{\lambda}$ in each iteration step of Newton's method satisfy

$$\sum_{k=1}^{4} [G_{k+l}(\boldsymbol{\lambda}) - m_l G_k(\boldsymbol{\lambda})]\, \Delta\lambda_k = G_l(\boldsymbol{\lambda}) - m_l G_0(\boldsymbol{\lambda}) \qquad (l = 1, \cdots, 4), \quad \text{(A13)}$$

where Eq. (A10) is used. For $l \le 3$, $G_l(\boldsymbol{\lambda})$ is computed by numerical integration of Eq. (A9), whereas $G_l(\boldsymbol{\lambda})$ for $l \ge 4$ is computed by using Eq. (A11). Substitution of the solution $\boldsymbol{\lambda}$ and $\lambda_0 = \log G_0(\boldsymbol{\lambda})$ into Eq. (47) yields entropy $H[X]$. If we use normalized moments $\{\widetilde{m}_l\}_{l=1}^4$ instead of moments $\{m_l\}_{l=1}^4$ in Eqs. (A12) and (47), we obtain normalized entropy $\widetilde{H}[X]$.

Examples of maximum entropy distributions are illustrated in Fig. A1 along with the values of $\widetilde{H}[X]$, and examples of distributions with $\widetilde{m}_3 = 0$ are listed in Table A1. The second distribution in this table is the Gaussian distribution, which has the largest entropy under the constraint that moments up to the second order are given. The third and fourth distributions are the maximum entropy distributions under the constraint that $\int_{-\infty}^{\infty} |x|\, p(x)\, dx$ and $\int_{-\infty}^{\infty} \sqrt{|x|}\, p(x)\, dx$ are given, respectively, instead of the second-order moment.

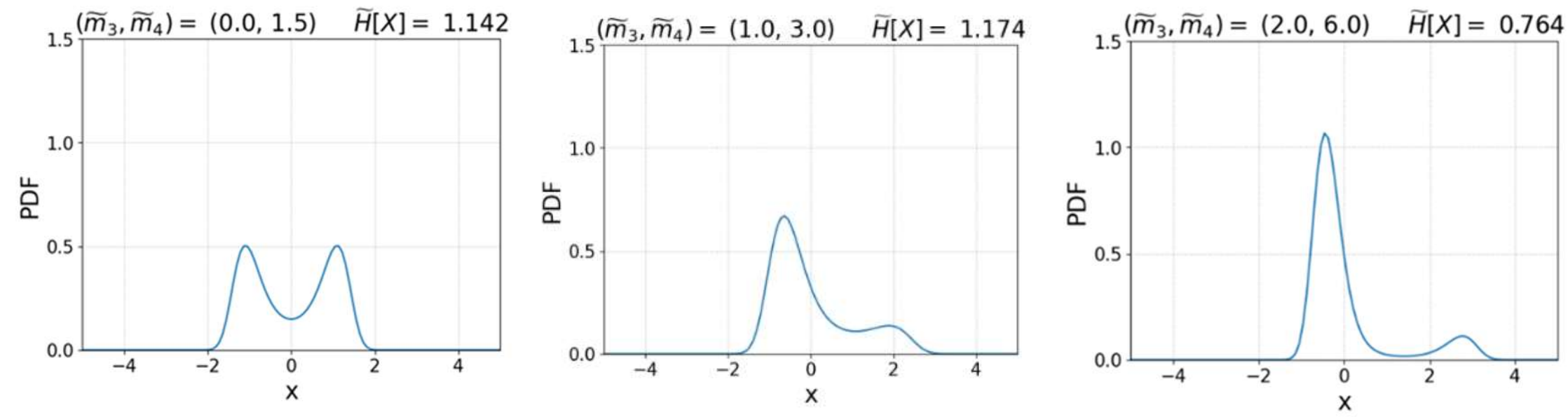

Fig. A1. Examples of maximum entropy distributions.

Table A1. Examples of distributions with $\widetilde{m}_3 = 0$.

| $(\widetilde{m}_3,\ \widetilde{m}_4)$ | PDF | $\widetilde{H}[X]$ |
|---|---|---|
| (0, 1) | $\frac{1}{2}[\delta(x+1)+\delta(x-1)]$ | $-\infty$ |
| (0, 3) | $\frac{1}{\sqrt{2\pi}}\exp\left(-\frac{x^2}{2}\right)$ | $\log\sqrt{2\pi}+\frac{1}{2}$ |
| (0, 6) | $\frac{1}{\sqrt{2}}\exp\left(-\sqrt{2}\lvert x\rvert\right)$ | $\log\sqrt{2}+1$ |
| (0, $\frac{126}{5}$) | $\sqrt{\frac{15}{2}}\exp\left(-\sqrt[4]{120}\sqrt{\lvert x\rvert}\right)$ | $\log\sqrt{\frac{2}{15}}+2$ |

The process of preparation of the lookup table is as follows. Because $\widetilde{m}_4 \geq |\widetilde{m}_3|^2 + 1$, we can start the computation mentioned above from the nearest grid points to the curve $\widetilde{m}_4 = |\widetilde{m}_3|^2 + 1$ with the first guess of $\boldsymbol{\lambda} = \mathbf{0}$. This first guess corresponds to a uniform distribution, and we need to change the range of integration in Eq. (A9) to a finite range to avoid divergence. After that, the convergent value of $\boldsymbol{\lambda}$ at the previous grid point adjacent to the current grid point is used as the first guess of the current grid point. This computation is conducted in a couple of directions in the $(|\widetilde{m}_3|, \widetilde{m}_4)$ plane to reduce convergence failure. A lookup table thus obtained is presented in Fig. 2a.

There are blank areas on the left of $\widetilde{m}_4 = 5\,|\widetilde{m}_3|^{2.5} + 3$ in this table, where Newton's method has failed to converge. We fill in the area on the left of this graph with mixture distributions. Along the $\widetilde{m}_4$-axis, the mixture distributions for $3 < \widetilde{m_4} \leq 6$ are constructed by linear interpolation of the second and third PDFs in Table A1, and those for $6 < \widetilde{m_4} \leq 9$ are constructed by linear interpolation of the third and fourth PDFs. The mixture distributions between the $\widetilde{m}_4$-axis and the graph of $\widetilde{m}_4 = 5\,|\widetilde{m}_3|^{2.5} + 3$ are constructed by linear interpolation of the mixture distribution on the $\widetilde{m}_4$-axis and the maximum entropy distribution

on that graph along the direction of $|\tilde{m}_3|$-axis. We compute the entropy of mixture distributions by numerical integration. The lookup table thus obtained is presented in Fig. 2b.

## Appendix 3. Proof of $\tilde{\boldsymbol{m}}_4 \geq |\tilde{\boldsymbol{m}}_3|^2 + 1$

Let $p(x)$ be a probability density function of a random variable $x$ satisfying the following equations:

$$\tilde{m}_1 = \int_{-\infty}^{\infty} xp(x)dx = 0, \tag{A14}$$

$$\tilde{m}_2 = \int_{-\infty}^{\infty} x^2 p(x)dx = 1. \tag{A15}$$

Then, the square of normalized third-order moment $\tilde{m}_3$ satisfies

$$\begin{aligned} {\tilde{m}_3}^2 &= \left[\int_{-\infty}^{\infty} x^3 p(x)dx\right]^2 = \left[\int_{-\infty}^{\infty} (x^3 - ax)p(x)dx\right]^2 \\ &\leq \int_{-\infty}^{\infty} x^2 p(x)dx \cdot \int_{-\infty}^{\infty} (x^2 - a)^2 p(x)dx = \int_{-\infty}^{\infty} (x^2 - a)^2 p(x)dx, \end{aligned} \tag{A16}$$

where $a$ is an arbitrary real constant, and the Cauchy-Schwarz inequality is used in the second line. From this result, we can derive the following inequality for normalized fourth-order moment $\tilde{m}_4$:

$$\tilde{m}_4 \geq |\tilde{m}_3|^2 - a^2 + 2a. \tag{$A17$}$$

The strongest inequality

$$\tilde{m}_4 \geq |\tilde{m}_3|^2 + 1 \tag{$A18$}$$

is obtained by setting $a$ to 1.